\documentclass{aa}
\usepackage{graphicx}
\usepackage{amsmath}
\usepackage{amssymb}
\usepackage{natbib}
\bibpunct{(}{)}{;}{a}{}{,}

\newcommand{\f}{f^{(\nu)}}
\newcommand{\be}{\begin{equation}}
\newcommand{\ee}{\end{equation}}

\usepackage{txfonts}

\begin{document}
   \title{A family of models of partially relaxed stellar systems}

   \subtitle{II. Comparison with the products of collisionless collapse}

   \author{M. Trenti\inst{1}
          \and
           G. Bertin\inst{2}
          \and
          T. S. van Albada\inst{3}
          }

   \offprints{m.trenti@sns.it}

     \institute{Scuola Normale Superiore, Piazza dei Cavalieri 7,
              I-56126 Pisa, Italy \\
           \email{m.trenti@sns.it}
           \and
       Dipartimento di Fisica, Universit\`{a} di Milano, via
Celoria 16, I-20133 Milano, Italy \\
 \email{Giuseppe.Bertin@unimi.it}
       \and
Kapteyn Astronomical Institute, Postbus 800, 9700 AV Groningen,
The Netherlands \\ \email{albada@astro.rug.nl}
     }

   \date{Received 21/07/04 ; accepted 15/11/04}

   \abstract{

N-body simulations of collisionless collapse have offered
important clues to the construction of realistic stellar dynamical
models of elliptical galaxies. Understanding this idealized and
relatively simple process, by which stellar systems can reach
partially relaxed equilibrium configurations (characterized by
isotropic central regions and radially anisotropic envelopes), is
a prerequisite to more ambitious attempts at constructing
physically justified models of elliptical galaxies in which the
problem of galaxy formation is set in the generally accepted
cosmological context of hierarchical clustering.

In a previous paper, we have discussed the dynamical properties of a
family of models of partially relaxed stellar systems (the $f^{(\nu)}$
models), designed to incorporate the qualitative properties of the
products of collisionless collapse at small and at large radii. Here
we revisit the problem of incomplete violent relaxation, by making a
direct comparison between the detailed properties of such family of
models and those of the products of collisionless collapse found in
$N$-body simulations that we have run for the purpose. Surprisingly,
the models thus identified are able to match the simulated density
distributions over nine orders of magnitude and also to provide an
excellent fit to the anisotropy profiles and a good representation of
the overall structure in phase space. The end-products of the
simulations and the best-fitting models turn out to be characterized
by a level of pressure anisotropy close to the threshold for the onset
of the radial-orbit instability. The conservation of $Q$, a third
quantity that is argued to be approximately conserved in addition to
total energy and total number of particles as a basis for the
construction of the $\f$ family, is discussed and tested numerically.

  \keywords{stellar dynamics --- galaxies: evolution --- galaxies:
formation --- galaxies: kinematics and dynamics --- galaxies:
structure}
   }

   \maketitle
%

\section{Introduction}

The collapse of a dynamically cold cloud of stars can lead to the
formation of realistic stellar systems, with projected density
profiles well represented by the $R^{1/4}$ law \citep{van82}. The
theoretical framework for the mechanism of incomplete violent
relaxation that governs this process of structure formation was
proposed by \citet{lyn67}, who argued that fast fluctuations of
the potential during collapse would lead to the formation of a
well-relaxed isotropic core, embedded in a radially anisotropic,
partially relaxed halo. This general picture served as a physical
justification for the construction of the so-called $f_{\infty}$
models, which indeed recovered the $R^{1/4}$ law and, suitably
extended to the case of two-component systems (to account for the
coexistence of luminous and dark matter), led to a number of
interesting applications to the observations \citep[see][ and
references therein]{ber84,ber93}.

An attempt at deriving the relevant distribution function directly
from the statistical mechanics of incomplete violent relaxation
suggested that, in addition to the $f_{\infty}$ models, one could
consider alternative models, called the $\f$ models \citep{sti87},
with similar overall characteristics. The key ingredient for the
construction of the $\f$ distribution function is the conjecture that
a \emph{third} quantity $Q$, in addition to the total mass $M$ and the
total energy $E_{tot}$, is {\it approximately} conserved during the
process of collisionless collapse (of course, we are referring to
systems characterized by vanishing total angular momentum,
$J_{tot}=0$). This quantity is introduced to model the process of {\it
incomplete} violent relaxation, ensuring a radially biased pressure
tensor and a $1/r^4$ density profile in the outer parts of the
system. Because of their relatively straightforward derivation from
the Boltzmann entropy, these models were revisited recently
\citep{ber03} and used to demonstrate the onset of the gravothermal
catastrophe \citep{lyn68} for such a one-parameter sequence (at fixed
$\nu$) of anisotropic equilibria; a preliminary inspection of the
general characteristics of the $\f$ models then convinced us that,
with significant advantage over the $f_{\infty}$ models, they might
also serve as a good framework to interpret the results of simulations
of collisionless collapse not only qualitatively, but also in {\it
quantitative detail}. Therefore, we proceeded to examine their
intrinsic properties systematically (\citealt{tre04}; hereafter Paper
I), and we will take advantage of that work for the study
presented in this paper.

In this article we describe the results of a relatively wide set
of numerical simulations of collisionless collapse, aimed at
studying the phase space evolution and settling of the system
during violent relaxation, and we then compare in detail the
properties of the quasi-equilibrium end-products thus obtained
with those of the $\f$ models. In particular, we discuss the role
played by the initial conditions and find that a certain degree of
clumpiness is required for an efficient mixing in the
single-particle angular momentum distribution; this form of
relaxation turns out to be crucial for a good match with the $\f$
family of models. The $Q$ conservation is then studied directly by
looking at its time evolution. For a
significant range of collapse factors, 
as determined by the initial values of the virial ratio $u = (2K/|W|)_{t
= 0}$, an approximate conservation is indeed observed. The
end-products (and thus the best-fitting models) tend to be
characterized by a value of the global anisotropy parameter close to
marginal stability with respect to the radial orbit instability
\citep{pol81}.

The paper is organized as follows. After introducing our basic models
and notation (Sect.~\ref{sec:units}), in Sect.~\ref{sec:code} we start
by reviewing the choice of the numerical code and we then check the
results obtained in some test runs against the tree code of
\citet{deh00}. In Sect.~\ref{sec:ic} we discuss the initial conditions
adopted for the simulations of collisionless collapse, with special
attention to the issue of clumpiness in phase space.  In
Sect.~\ref{sec:end_prod} we characterize the end-products of the
simulations in terms of a few key indicators (i.e., central
concentration, global anisotropy, density and anisotropy profiles,
deviations from spherical symmetry) and describe their dependence on
the initial conditions. In Sect.~\ref{sec:Q} we examine the hypothesis
of the approximate conservation of $Q$. We then move, in
Sect.~\ref{sec:fit}, to the comparison of the end-products of the
simulations with the $\f$ models (in terms of density and anisotropy
profiles and directly in phase space). In Sect.~\ref{sec:con}, we draw
the main conclusions from this study.  Finally, in the Appendix we
provide additional comments on the issue of clumpiness in phase space.

\section{$\f$ models, units, and notation}\label{sec:units}

In general, we will keep the same notation as in Paper I. We recall
that the relevant distribution function is obtained by extremizing the
Boltzmann entropy $S=-\int f \log f d^3x d^3w$ at fixed total mass
$M=\int f d^3x d^3w$, total energy $E_{tot}=1/3 \int E d^3x d^3w$, and
$Q = \int J^{\nu} |E|^{-3 \nu/4} f d^3x d^3w$. Here $E$ and $J$ denote
single-star specific energy and angular momentum, while $\vec{x}$ and
$\vec{w}$ denote positions and velocities respectively.
This leads to the function $\f = A \exp {[- a E - d
(J^2/|E|^{3/2})^{\nu/2}]}$, where $a$, $A$, $d$, and $\nu$ are
positive constants. The $\f$ models are then constructed by solving
the Poisson equation for the unknown potential $\Phi(r)$
numerically. At fixed value of $\nu$, one may think of the free
constants as providing two dimensional scales (for example, $M$ and
$E_{tot}$) and one dimensionless parameter, such as $\Psi \equiv -a
\Phi(0)$, the central depth of the dimensionless potential well. By
$(1;5)$ $\f$ model we will denote the model of the $\f$ family with
$\nu = 1$ and $\Psi = 5$.

The $\f$ models represent equilibrium configurations designed to
describe the products of incomplete violent relaxation. They are
characterized by a density profile $\rho(r)$ falling off as $1/r^4$ at
large radii and as $1/r^2$ in the inner part of the system, outside a
central ``core''. The size of the core becomes smaller as the
concentration parameter $\Psi$ increases. On the large scale, apart
from such freedom in central concentration and core size, the shape of
the density profile is basically independent of the $(\nu;\Psi)$
parameters (see Fig.~3 in Paper I).  Interestingly, although this
feature had not been imposed at the beginning (when the function $\f$
is constructed), the projected density distribution of the $\f$ models
is typically well fitted, on the large scale, by the $R^{1/4}$ law;
residuals in the fit are reduced if one considers the generalized
$R^{1/n}$ law (with $n$ a free parameter; \citealt{ser68}), depending
on $\Psi$ (see Figs.~4-5 in Paper I).

In contrast with other approaches (e.g., see \citealt{osi79} and
\citealt{mer85c}) where the anisotropy profile is assigned \emph{a
priori}, in the $\f$ models the velocity dispersion anisotropy profile
$\alpha(r)$, defined as $\alpha(r) = 2 - (\langle w^2_{\theta}\rangle
+ \langle w^2_{\phi} \rangle)/\langle w^2_r \rangle$, must be computed
\emph{a posteriori} and its shape depends on $\nu$ and $\Psi$ (see
Fig.~6 in Paper I). The structure of the distribution function only
guarantees that the models match the asymptotic requirements suggested
by the picture of incomplete violent relaxation, i.e. at large radii,
where the pressure is radial, and in the central regions, where the
pressure is isotropic. The global anisotropy, measured by the quantity
$2K_r/K_T$, i.e. twice the ratio of the radial to the tangential
kinetic energy, depends on the choice of $(\nu;\Psi)$ and correlates
with the central concentration (e.g., see Fig.~7 in Paper I). Models
with $\Psi \lesssim 4$ are characterized by an excessive degree of
radial anisotropy (i.e. $2K_r/K_T \gtrsim 1.7$), and are thus
unstable.

The physical system of units adopted in this paper is defined by
$10~kpc$ for length, $10^{11}~M_{\odot}$ for mass, and $10^8~yr$ for
time. In this system, natural for studies on the galactic scales,
velocities are measured in units of $\approx 97.8~km/s$ and the value
of the gravitational constant $G$ is $4.4971$.

The majority of simulations consists of runs starting from $20$
cold clumps of $16~kpc$ radius in a sphere of $40~kpc$ radius,
with $u=(2K/|W|)_{t = 0}$ in the range $0.05$-$0.25$. After the
collapse the system has a half-mass radius around $8~kpc$. The
total mass of the system is $2 \times 10^{11}~M_{\odot}$. The
dynamical time, which we define as $t_d = GM^{5/2}/(- 2
E_{tot})^{3/2}$, is therefore typically $\approx 1.2 \times
10^8~yr$, i.e. $1.2$ in our units. As a result, when we stop the
simulation at time $80$, the system has evolved for several tens
of dynamical times. In any case, we should recall that the results
obtained are scale-free, that is they can be rescaled to other
choices of mass and radius if so desired.

\section{The code}\label{sec:code}

Direct N-body simulations of self-gravitating stellar systems
require huge amounts of computing time because of the $N^2$
scaling of the code complexity with the number of particles
employed. To model the evolution of collisionless systems, several
algorithms have been developed that treat the gravitational
interactions approximately, with a lower complexity. In principle,
to study the process of collisionless collapse we have two
options: either a tree code \citep[e.g.,][]{bar86} or a
particle-mesh like algorithm \citep[e.g.,
see][]{van82,mcg84,her92}. Tree codes are intrinsically slower,
scaling as $N\log{N}$, with respect to particle-mesh Poisson
solvers, for which the complexity is linear in $N$, but the latter
have the disadvantage of a lower spatial resolution, being limited
by the size of the grid used.

In this paper we are interested in the large scale structure of the
end-products of collisionless collapse, for systems that do not
exhibit large deviations from spherical symmetry. The natural choice
thus appears to be that of a particle-mesh code, based on a spherical
grid and an expansion in spherical harmonics.

The code used in the present study is thus a new version of the
\citet{van82} code. The relevant changes introduced are briefly
described below, in Sect.~\ref{sec:code_des}. For completeness, we
have also run (see Sect.~\ref{sec:falcon}) a number of comparison
simulations with the fast code developed by \citet{deh00}.

\subsection{An improved particle-mesh code} \label{sec:code_des}

The key feature of the \citet{van82} code is the solution of the
Poisson equation $\nabla^2 \Phi = 4 \pi G \rho$, which relates the
mean potential $\Phi$ of the system to the mass density $\rho$, by
means of Fourier techniques. Once the potential has been computed
by expanding the density in spherical harmonics, the acceleration
is obtained by numerical differentiation, and the particles are
advanced by a fixed time step, using a leap-frog scheme.

At variance with the original implementation, to preserve accuracy
and to avoid systematic errors, we decided to drop the angular
grid and to treat the angular dependence of the force exactly, in
terms of the single-particle Legendre polynomials (for further
details, see \citealt{tren04}). Basically, this choice changes the
code in the direction of the code of \citet{mcg84} and of the
self-consistent field code of \citet{her92}. We preferred to keep
the radial grid because of its flexibility with respect to the
density profile, especially under conditions of rapid evolution
(we use a subroutine to generate a grid containing a fixed
fraction of the total mass in each shell).

The density is assigned to the radial grid by means of a cloud-in-cell
scheme with a linear kernel, i.e. a particle contributes to the
density of the two closest cells with a weight depending linearly on
the distance from the center of the cell considered. The same kernel
is then used to assign the force from the grid to the particle. The
time step is adaptively chosen in such a way that particles are not
allowed to cross more than one radial cell during one step.

\begin{figure}
\resizebox{124pt}{!}{\includegraphics{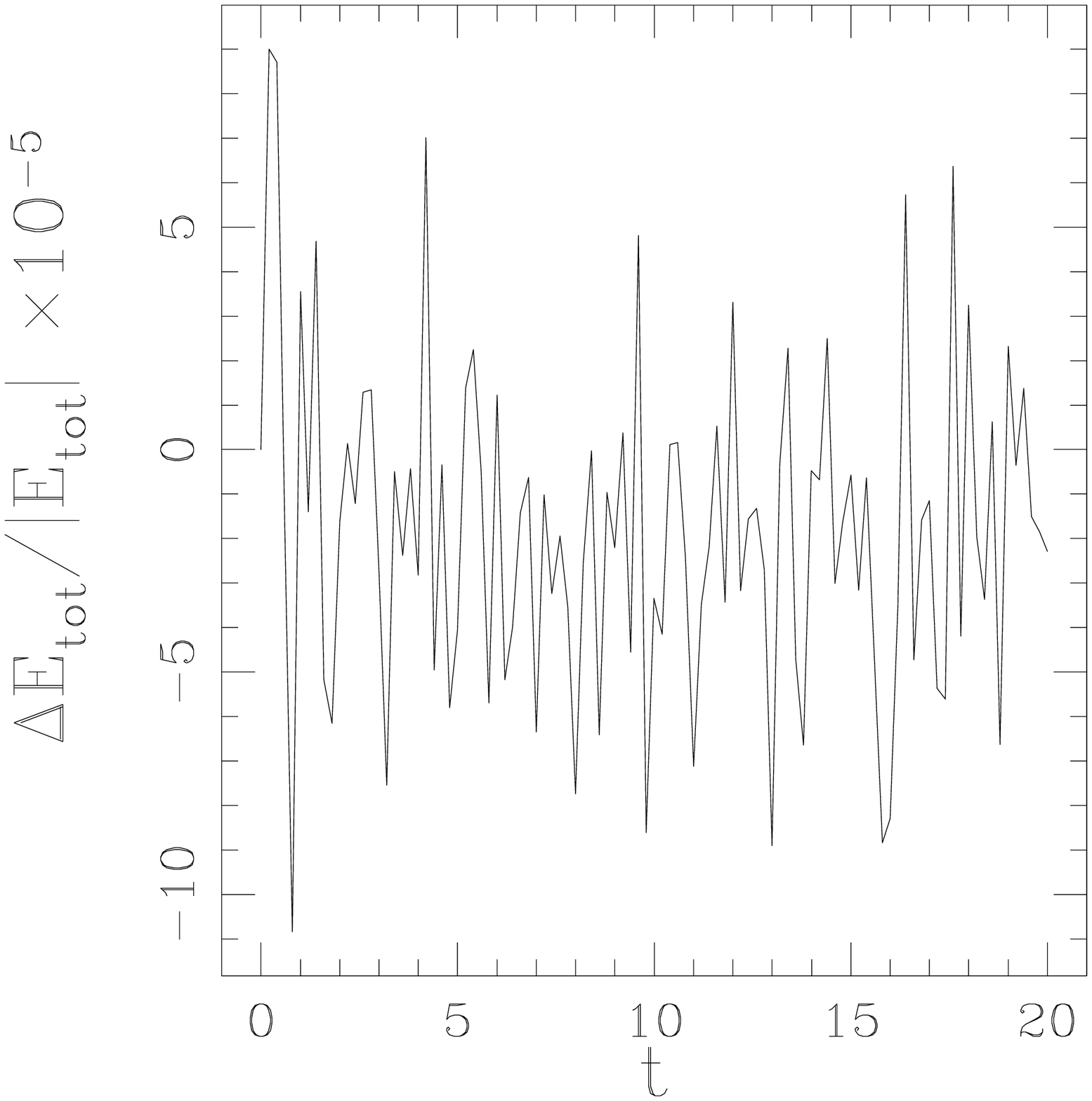}}
\resizebox{124pt}{!}{\includegraphics{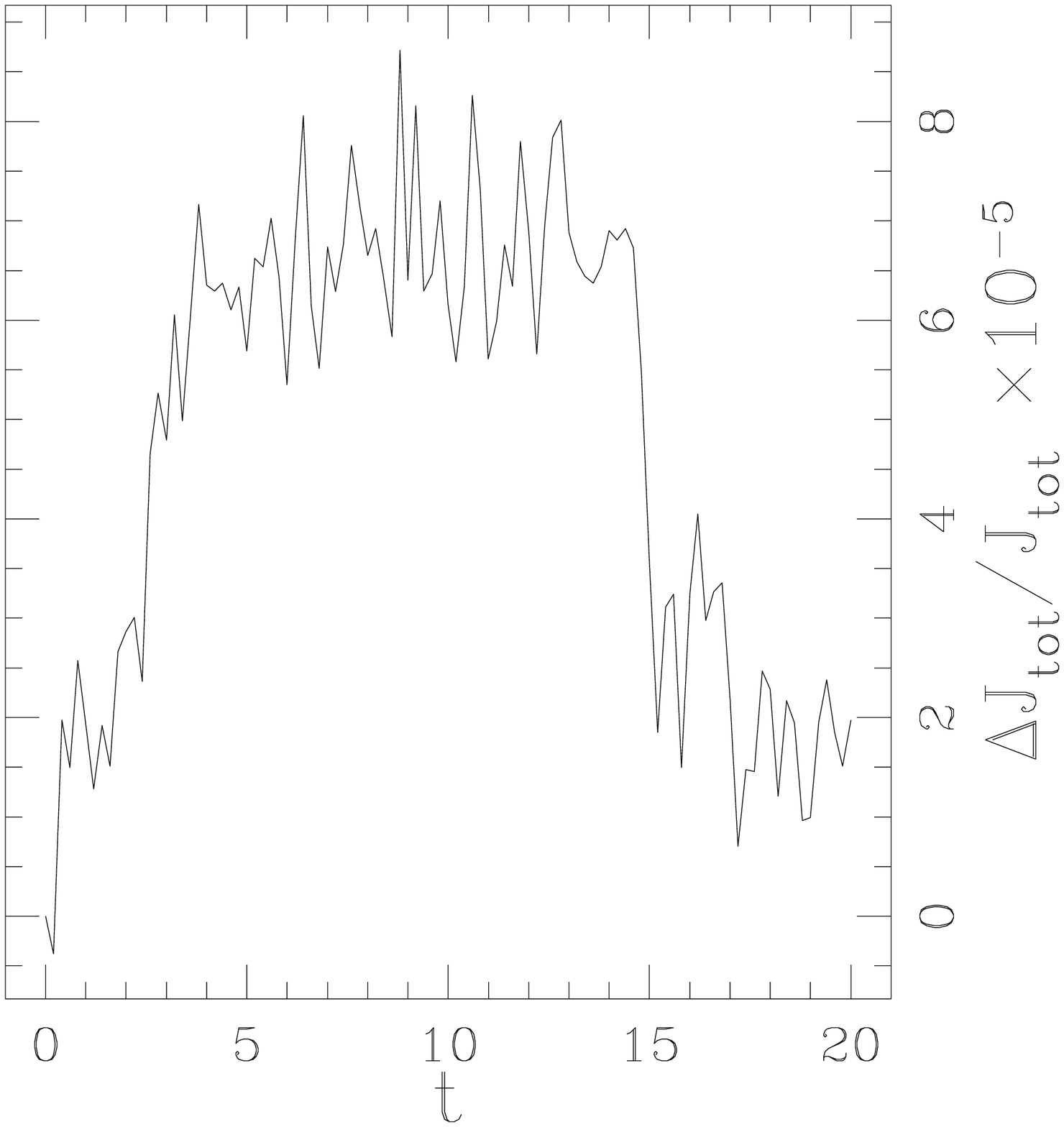}}
 \caption{Total
  energy conservation (left) and angular momentum (right) for a
  simulated $(1;5)$ $\f$ equilibrium model with $4\times 10^{5}$
  particles. The mass of the system is $M=1$, the total energy
  $E_{tot}=-1.73$, and the half-mass radius $r_M=0.5$. The magnitude
  of the total angular momentum associated with the initial conditions
  corresponds to a value of the dimensionless rotation parameter
  $\lambda = J_{tot} |E_{tot}|^{1/2}/(G M^{5/2}) \approx 10^{-4}$. We
  recall that time is given in units of $10^{8}$ years, which, in this
  case, corresponds to $\approx 1.4~t_d$, with $t_d =
  GM^{5/2}/(-2E_{tot})^{3/2}$.}
  \label{fig:de_Psi5}
\end{figure}

The code has been tested extensively, in terms of its accuracy in
conserving total energy and total angular momentum for equilibrium
and non-equilibrium initial conditions and in conserving
single-particle energy and angular momentum for runs of
spherically symmetric equilibrium models. The modified Poisson
solver scheme combined with the adaptive grid ensures a
significantly better accuracy with respect to the original
implementation. The typical total energy and total angular
momentum conservation for a run with $10^5$ particles is of the
order of $10^{-5}$ per dynamical time in quasi-stationary
configurations (see Fig.~\ref{fig:de_Psi5}).
\subsection{Comparison with Dehnen's code}\label{sec:falcon}

As a further test, we have also run a few test simulations by
comparing, under identical conditions, the performance of our code
to that of the fast tree code \emph{GyrFalcON}
\citep{deh00,deh02}, within the {\it NEMO} distribution
\citep{teuben}. In such tests, we adopt the following procedure.
We first generate the initial conditions in the physical units
used by our code (see Sect.~\ref{sec:units}) and we run the
simulation. We then convert the initial conditions to the natural
units defined by \citet{heggie} and used in Dehnen's code.
Finally, we run the simulation within the {\it NEMO} environment.
The quality of the integration is checked with the standard {\it
NEMO} tools of analysis. At the end of the simulation, a
``snapshot" of the system is exported and converted back to our
units, in such a way that it can be processed by the same
diagnostics used for the particle-mesh code.

The initial conditions for these runs have been chosen in order to
be representative of the sample investigated in this paper; they
are described in Table~\ref{tab:ic} ($C4.1$ and $C4.3$ entries),
with the properties of the final equilibrium state listed in
Table~\ref{tab:main}.


For the runs with Dehnen's tree code we adopted the following choice
of integration parameters: tolerance parameter $\theta=0.5$ (standard
choice $0.6$) to improve accuracy in the calculation of forces;
softening length $\epsilon=0.01$ (in natural units; standard choice
$0.05$) to increase central resolution; minimum allowed time step
$1/2^8$ (i.e. $\approx 724$ steps per dynamical time). With this
choice of integration parameters, the energy and angular momentum
conservation is very good: in one dynamical time $t_d$, the relative
changes are $\Delta E_{tot}/E_{tot} < 10^{-5}$ and $\Delta
J_{tot}/J_{tot} < 10^{-4}$.

The required CPU time to complete the simulation is marginally higher
than with our code, which, however, has not yet been optimized for
speed.

\begin{figure}
  \resizebox{\hsize}{!}{\includegraphics{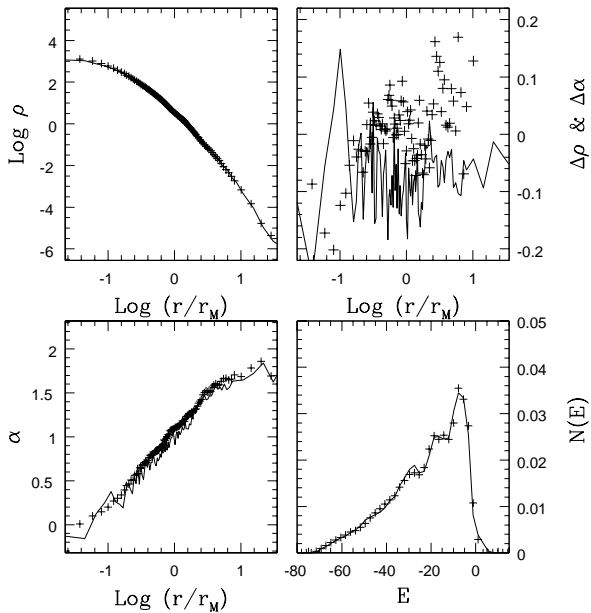}}
\caption{ Comparison between our code and \emph{GyrFalcON}
\citep{deh00} for the run $C4.1$. Final density (upper left) and
anisotropy (bottom left) profiles, and single-particle energy
distribution (bottom right) for a run starting from $10^5$
particles in $10$ cold clumps with $u=0.15$ (see
Tables~\ref{tab:ic} and \ref{tab:main} for further details about
the simulation); here the solid lines give the results obtained
with \emph{GyrFalcON}, while the crosses refer to our code. In the
upper right panel we plot the differences in the density profile
$\Delta \rho = 2
(\rho_{falcon}-\rho_{pm})/(\rho_{falcon}+\rho_{pm})$ (crosses) and
in the anisotropy profile $\Delta \alpha =
\alpha_{falcon}-\alpha_{pm}$ (solid line).  }
  \label{fig:comp}
\end{figure}

As desired, for these runs we find a substantial similarity in the
properties of the end-products obtained by the two different
methods of integration (see Fig.~\ref{fig:comp}). To be sure,
small differences naturally arise, as expected. The main
systematic difference is in the degree of anisotropy
characterizing the end-products of the simulations. In fact, the
output from the tree code is slightly more isotropic: the final
global anisotropy, measured by $2K_r/K_T$, is up to $7\%$ lower,
with a slight outward shift of the anisotropy profile,
corresponding to a more efficient core relaxation. This might be
related to the residual collisionality present in the tree code.

\section{Choice of initial conditions}\label{sec:ic}

If the initial conditions are not too artificial, during the
process of collisionless collapse, violent relaxation can take
place, with significant mixing in phase space, and wipe out much
of the details that characterize the initial conditions. In
reality, violent relaxation is incomplete. Therefore, the final
state is that of an approximate dynamical equilibrium
characterized by an anisotropic distribution function, different
from a Maxwellian (which would correspond to thermodynamic
equilibrium). Because of such incomplete relaxation, the
end-products of the simulations do conserve some memory of the
initial state.

\subsection{Uniform initial conditions, clumpy initial conditions,
and the cosmological framework} \label{sec:ic_cosmo}

Some of the papers addressing the problem of collisionless collapse
start from ``uniform" initial conditions in position and velocity
space. For example \citet{agu90} assume an initial $1/r$ density
profile and then explore the way the collapse proceeds, by varying, in
addition to the initial virial ratio $u=(2K/|W|)_{t = 0}$, the shape
of the initial density profile (by shrinking the system along one
axis) and the amount of rotation. \citet{udr93} starts from uniform
cold spheres, and also varies, in addition to the above-mentioned
parameters, the initial anisotropy content $2K_r/K_T$. Recently,
\citet{boi02}, starting from cold uniform spheres or spheroids, focus
on the effects introduced by the number of particles used in the
simulation. 

A few earlier investigations \citep{van82,mcg84,may84,lon91} compared
``clumpy'' to ``uniform" (or ``homogeneous'') initial conditions,
showing that clumpy initial conditions lead to end-states with
projected density distributions well fitted by the $R^{1/4}$ law
(although \citealt{agu90} point out that, for very small values of
$u$, the $R^{1/4}$ law is approximately recovered even for homogeneous
initial conditions). [\citet{udr93} argues that starting from a
multi-component initial mass spectrum for the simulation particle
distribution can be be an alternative way to represent a clumpy
initial density configuration. However, the introduction of simulation
particles as massive as to be representative of clumps, would
introduce effects of dynamical friction that {\it per se} would go
beyond the picture of collisionless violent relaxation.] 

Recently \citet{roy04} studied the outcome of violent collapse
starting from an initial uniform background with the possible addition
of small clumps of stars. Although their clumpy initial conditions are
rather different from those considered in the present paper, they also
noted that clumpy simulations lead to steep density profiles with
small cores.

As will also be demonstrated later on (see Sect.~\ref{sec:jscatter}),
the key point that distinguishes clumpy from uniform initial
conditions is that, in general, only the former allow for significant
mixing in phase space, thus making it possible for violent relaxation
to proceed properly. Thus in this paper we will focus on simulations
starting from clumpy configurations. As discussed in the next
Subsection, for the present study we do not require that our initial
clumps be in internal dynamical equilibrium, since their purpose is to
avoid excessive homogeneity in the $(E,J^2)$ phase space (see also
Appendix). In particular, the clumps are \emph{not} intended to be a
realistic representation of possible conditions at a given epoch in
the past. In fact, the effects of violent relaxation become important
in a few dynamical times, independently of the precise epoch when the
process is imagined to occur.

To be sure, to identify a realistic set of initial conditions, one
should consider a satisfactory cosmological framework. We plan to do
this in future investigations, because this would lead us well beyond
the scope of the present paper. In this respect, the use of clumps is
already one important step forward with respect to the use of
homogeneous initial conditions. Eventually, we should devise a method
for determining a ``spectrum'' of clumps with properties compatible
with the expectations of current cosmological scenarios \citep[see
also][ and further discussion in the Appendix]{kat91}. For the moment,
we are satisfied with identifying the initial conditions under which
sufficient mixing in phase space is guaranteed. Note that
cosmologically oriented simulations are centered on the clustering and
growth of dark matter halos, while in this paper, given our focus on
the $R^{1/4}$ law and on the deviations from it, we have in mind
mostly luminous matter.

\subsection{Setting up clumpy initial conditions}

In a clumpy initial state the $N$ particles are grouped in $N_C$
spherical clumps, each of them containing $N_i$ stars, so that
$N=\sum_{i=1}^{N_C}N_{i}$, with $\langle N_i \rangle = N/N_C$.  Within
each clump the star distribution is homogeneous. The centers of mass
of the clumps are distributed uniformly inside a sphere of radius $R$,
which defines the size of the system at the beginning of the
simulation. The clump radius is $R_C$, with $R_C<R$, but such that
$N_C \times R_C^3 > R^3$ (this condition ensures that the sphere of
radius $R$ is well filled by stars).  The initial kinetic energy may
be associated with the ordered motion of the center of mass of each
clump (this is our default choice for the simulations of type $C$
described below; in this case the velocity is assigned by drawing from
an isotropic distribution) or with the random motions of the stars
within the clump (in these cases we add a subscript \emph{h} to the
simulation label; here the center of mass of each clump is taken to be
at rest). In general, with this choice of initial conditions the
clumps are not in internal dynamical equilibrium. We note that when
the number of clumps used is low, the initial configuration may
deviate significantly from spherical symmetry (with projected shapes
up to those of an $E3$ galaxy). Formally the limits $N_C
\longrightarrow N$ and $N_C \longrightarrow 1$ both lead to the case
of homogeneous initial conditions.

\begin{figure}
  \resizebox{\hsize}{!}{\includegraphics{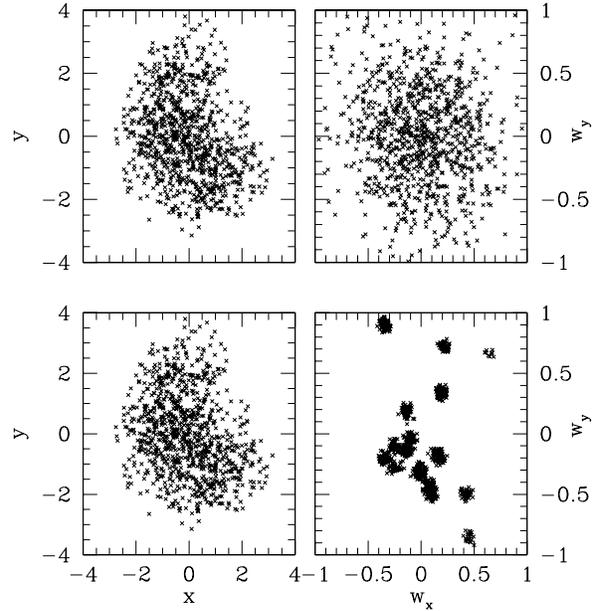}} \caption{
Typical projected distributions in position (left) and velocity
(right) space for hot (upper panels, run $C4.1_h$) and cold (lower
panels, run $C4.1$) clumpy initial conditions.}
  \label{fig:ic_cl}
\end{figure}

In the case of homogeneous initial conditions (simulations of type $U$
and $S$), which we run for comparison, we employed two kinds of
distribution: (1) a constant density within a sphere of radius $R$;
(2) a symmetrized version of a given clumpy configuration (simulations
of type $S$). The symmetrization process in (2) is performed by
accepting the radius and the magnitude of the velocity of each
particle, following the procedure for initial clumpy conditions, but
by redistributing uniformly the angular variables.

In principle, we have a wide parameter space to explore, because
we have to deal with the initial virial ratio $u$, the number and
size of the clumps, the cold/hot choice for the initial kinetic
energy distribution (and the intermediate range of possibilities),
the spatial distribution of the centers of mass of the clumps and
of the stars within each clump. As noted earlier \citep[e.g.,
see][]{lon91}, we anticipate that the main controlling physical
parameter is the initial virial ratio.

Table~\ref{tab:ic} lists for each simulation the following
information: the number $N$ of particles used, the number of clumps
$N_C$, the initial virial ratio $u$, the initial values of the shape
parameters $\epsilon_0$, $\eta_0$ (based on the length of the axes of
the homogeneous ellipsoid associated with the inertia tensor, taken to
be in the order $a \ge b \ge c$, so that $\eta = c/a$ and $\epsilon =
b/a$; the inertia tensor is referred to the particles within a sphere
of radius $3r_M$), and the initial concentration
${C_{\rho}}_0=\left(\rho(0)/\rho(r_M)\right)_{t=0}$. As a summary for
the notation used, we note the following. We have divided the set of
clumpy simulations in five subsets, from $C1$ to $C5$. The simulations
belonging to $C1$ start with $10^5$ particles in $10$ clumps, the
positions of which are fixed. The $C2$ series is a high resolution ($8
\times 10^5$ particles) version of $C1$, but uses instead 20
clumps. In the $C3$ (high resolution, $8 \times 10^5$ particles) and
$C4$ ($10^5$ particles) series we use different seeds for the initial
positions of the clumps and we also change other parameters as
described in Table~\ref{tab:ic}.  Runs $CV5.1$ and $CP5.2^*$ are test
runs especially performed to clarify some issues related to clumpiness
(see Appendix). $CV5.1$ has clumpy conditions in velocity space as in
run $C4.1$, but uniform homogeneous conditions in position space; in
turn, $CP5.2^*$ has a clumpy configuration in position space ($40$
clumps of $6~kpc$, with a filling factor $N_C \times R_C^3 / R^3 =
0.135$) and uniform conditions in velocity space. Runs $U$ refer to
uniform homogeneous spheres (here the seed for the random numbers is
not relevant given the high symmetry of the configuration) and the $S$
series refers to the symmetrized runs.

\begin{table}[!h]
\caption{Initial conditions for the simulations. After the simulation
  identifier the columns list the number of particles $N$, the number
  of clumps $N_C$, the initial virial ratio $u$, the initial values of
  the shape parameters $\epsilon_0$ and $\eta_0$, and the initial
  concentration $C_{\rho 0}$. For the exact definitions and for the
  general characteristics of the groups ($C1$ to $C4$, $U$ and $S$),
  see description in Sect.~4.2. Four simulations, marked by a $^*$,
  have been carried out with {\it GyrFalcOn} \citep{deh00}; two of
  them ($C4.1^*$ and $C4.3^*$) start from the same initial conditions
  as for $C4.1$ and $C4.3$ respectively, while $C4.4^*$ and $CP5.2^*$
  are characterized by small clumps (with radius $2.8~kpc$ and $6~kpc$
  respectively, distributed in a sphere of radius $40~kpc$).}
\begin{center}
\begin{tabular}{ccccccc}
\hline \hline &  $N$ & $N_C$ & $u$ & $\epsilon_0 $ & $\eta_0$ &
${C_{\rho}}_0$\\ \hline

$C1.1$ &$10^5$ & 10  & 0.275 & 0.83 & 0.70 &3.0 \\
$C1.2$ &$10^5$ &10  & 0.25  & 0.83 & 0.70 & 3.0\\
$C1.3$ &$10^5$ &10  & 0.225 & 0.83 & 0.70 & 3.0\\
$C1.4$ &$10^5$ &10  & 0.20  & 0.83 & 0.70 & 3.0\\
$C1.5$ &$10^5$ &10  & 0.175 & 0.83 & 0.70 & 3.0\\
$C1.6$ &$10^5$ &10  & 0.15  & 0.83 & 0.70 & 3.0\\
$C1.7$ &$10^5$ &10  & 0.125 & 0.83 & 0.70 & 3.0\\
$C1.8$ &$10^5$ &10  & 0.1   & 0.83 & 0.70 & 3.0\\
$C1.9$ &$10^5$ &10  & 0.075 & 0.83 & 0.70 & 3.0\\
$C1.10$ &$10^5$ &10  & 0.05  & 0.83 & 0.70& 3.0\\
\hline
$C2.1$ &$8\cdot 10^5$ & 20 & 0.23 & 0.93 & 0.73 &2.8\\
$C2.2$ &$8\cdot 10^5$ & 20 & 0.17 & 0.93 & 0.73 &2.8\\
$C2.3$ &$8\cdot 10^5$ & 20 & 0.12 & 0.93 & 0.73 &2.8\\
$C2.4$ &$8\cdot 10^5$ & 20 & 0.06 & 0.93 & 0.73 &2.8\\

\hline $C3.1$ &$8\cdot 10^5$ & 20 & 0.08 & 0.95 & 0.91 & 2.2 \\
$C3.2$ &$8\cdot 10^5$ & 20 & 0.18 & 0.86 & 0.80 & 2.6 \\ $C3.3$
&$8\cdot 10^5$ & 20 & 0.15 & 0.84 & 0.70 & 3.1 \\ $C3.4$ &$8\cdot
10^5$ & 20 & 0.23 & 0.88 & 0.73 & 2.0 \\ $C3.5$ &$8\cdot 10^5$ &
20 & 0.15  & 0.95 & 0.88 & 3.7 \\ $C3.6$ &$8\cdot 10^5$ & 10 &
0.15  & 0.86 & 0.80 & 3.7 \\ \hline $C4.1$     & $10^5$&10  & 0.15
& 0.87 & 0.80 & 2.8 \\ $C4.1^*$   & $10^5$&10& 0.15 & 0.87 & 0.80
& 2.8  \\ $C4.1_{h}$ & $10^5$&10  & 0.15 & 0.87 & 0.80& 2.8 \\
$C4.2$     & $10^5$&20  & 0.25 & 0.75 & 0.63& 1.5 \\ $C4.3$     &
$10^5$&80  & 0.14 & 0.90 & 0.77& 1.9 \\ $C4.3^*$   & $10^5$&80  &
0.14 & 0.90 & 0.77& 1.9\\ $C4.4^*$   & $10^5$&80$^{+}$& 0.15 &
0.85 & 0.78& 2.0\\ $C4.5$     & $10^5$&400 & 0.23 & 0.99 & 0.95&
0.8 \\ $C4.5_{h}$ & $10^5$&400 & 0.23 & 0.99 & 0.95& 0.8 \\ \hline
$CV5.1$    & $10^5$& 10 & 0.23 & 1.00 & 1.00& 1.0 \\ $CP5.2^*$
& $10^5$& 40 & 0.15 & 0.81 & 0.78& 0.7 \\
\hline

$U6.1$     & $8\cdot 10^5$  & {N/A} & 0.10 & 1.00 & 1.00& 1.0\\
$U6.2$     & $8\cdot 10^5$  & {N/A} & 0.19 & 1.00 & 1.00& 1.0\\
$U6.3$     & $8\cdot 10^5$  & {N/A} & 0.29 & 1.00 & 1.00& 1.0\\
$U6.4$     & $8\cdot 10^5$  & {N/A} & 0.39 & 1.00 & 1.00& 1.0\\

\hline

$S4.2$ & $10^5$&{N/A} & 0.25 & 1.00 & 0.99 & 1.5\\
$S4.3$ & $10^5$ &{N/A}  & 0.15 & 1.00 & 1.00& 1.9 \\


\hline

\label{tab:ic}
\end{tabular}
\end{center}
\end{table}


\section{The products of collisionless collapse}\label{sec:end_prod}

\begin{table}[!h]
\caption{Final configurations for the simulations of collisionless
collapse listed in Table~\ref{tab:ic}.  The column entries are
described in Sect.~\ref{sec:end_prod}. Note that the anisotropy
profile in homogeneous simulations can be non-monotonic; this is
indicated by $^{\dagger}$.  All simulations of type $C1$ and $C2$
start from identical initial conditions within each series, except
for a constant scaling of velocities. The quantity $\Delta Q$ is
referred to $\nu=5/8$ in simulation $CV5.1$, to $\nu=1$ in
simulation $U6.1$ and to $\nu=3/4$ in simulation $U6.2$; this is
indicated by $^{\#}$.}
\begin{center}
\begin{tabular}{ccccccccc}
\hline \hline &  $\Delta
M$ & $\Delta Q$ & $C_{\rho}$ & $\kappa$ & $r_{\alpha}/r_M $ &
$\epsilon $  & $\eta$  \\ \hline

$C1.1$ & 0.00 & 0.13 & 570  & 1.61 & 1.02 & 0.91 & 0.73 \\
$C1.2$ & 0.002& 0.17 & 600  & 1.60 & 0.94 & 0.91 & 0.74 \\
$C1.3$ & 0.01 & 0.20 & 680  & 1.59 & 0.94 & 0.90 & 0.76 \\
$C1.4$ & 0.01 & 0.24 & 790  & 1.57 & 0.88 & 0.95 & 0.79 \\
$C1.5$ & 0.02 & 0.30 & 720  & 1.52 & 0.88 & 0.96 & 0.81 \\
$C1.6$ & 0.03 & 0.38 & 820  & 1.50 & 0.93 & 0.99 & 0.80 \\
$C1.7$ & 0.04 & 0.44 & 760  & 1.47 & 0.92 & 0.97 & 0.78 \\
$C1.8$ & 0.05 & 0.52 & 850  & 1.53 &  0.87  &  0.96 & 0.79 \\
$C1.9$ & 0.06 & 0.66 & 1130 & 1.67 &  0.75  &  0.97 & 0.75 \\
$C1.10$& 0.08 & 0.72 & 1090 & 1.74 &  0.79  &  0.94 & 0.69 \\
\hline
$C2.1$ & 0.01  & 0.13 & 110 & 1.52 & 1.49 & 0.87 & 0.78 \\
$C2.2$ & 0.02  & 0.25 & 160 & 1.62 & 1.24 & 0.88 & 0.78 \\
$C2.3$ & 0.03  & 0.4  & 270 & 1.70 & 0.83 & 0.81 & 0.69 \\
$C2.4$ & 0.07  & 0.5  & 520 & 1.76 & 0.74 & 0.81 & 0.63 \\

\hline
$C3.1$ & 0.003 & 0.47 & 1690 & 1.99 & 0.44 & 0.90 & 0.73 \\
$C3.2$ & 0.001 & 0.26 & 1250 & 1.85 & 0.55 & 0.93 & 0.70 \\
$C3.3$ & 0.04  & 0.57 & 430  & 1.60 & 1.34 & 0.92 & 0.71 \\
$C3.4$ & 0.02  & 0.23 & 500  & 1.65 & 1.15 & 0.93 & 0.81 \\
$C3.5$ & 0.005 & 0.24 & 950 & 1.73 & 0.57 & 0.96 & 0.72 \\ 
$C3.6$ & 0.005 & 0.27 & 690 & 1.79 & 0.75 & 0.80 & 0.77 \\ 
\hline
$C4.1$     & 0.005 & 0.27 & 440 & 1.77 & 0.83 & 0.80 & 0.73 \\
$C4.1^*$   & 0.01  & 0.27 & 360 & 1.68 & 0.97 & 0.89 & 0.74  \\
$C4.1_{h}$ & 0.00  & 0.18 & 240 & 1.86 & 0.51 & 0.86 & 0.83 \\
$C4.2$     & 0.12  & 0.10 & 160 & 1.40 & 1.65 & 0.90 & 0.78 \\
$C4.3$     & 0.10 &$<$0.01& 70  & 1.60 & 1.53 & 0.84 & 0.74 \\ 
$C4.3^*$   & 0.07 &  0.15 & 70  & 1.50 & 1.56 & 0.86 & 077  \\
$C4.4^*$   & 0.04  & 0.4  & 4000& 1.15 & 5.30  & 0.98 & 0.96\\
$C4.5$     & 0.125 & 0.02 & 20 & 1.20 & 1.74 & 0.96 & 0.95 \\
$C4.5_{h}$ & 0.10  & 0.05 & 15 & 1.16 & 1.58 & 0.99 & 0.97 \\
\hline
$CV5.1$    & 0.12  & 0.02$^{\#}$ & 90 & 1.55 & 1.50 & 0.91 & 0.75 \\ 
$CP5.2^*$    & 0.10  & 0.01 & 590 & 1.33 & 2.20 & 0.83 & 0.76\\
\hline

$U6.1$     & 0.33  & 0.29$^{\#}$ & 8 & 1.10 & $1.60^{\dagger}$ &
1.00 & 1.00 \\ $U6.2$     & 0.20  & 0.01$^{\#}$ & 6 & 1.10 &
$1.56^{\dagger}$ & 1.00 & 1.00 \\ $U6.3$     & 0.06  & 0.14 & 9 &
1.11 & $1.50^{\dagger}$ & 1.00 & 1.00 \\ $U6.4$     & 0.00  & 0.09
& 8 & 1.11 & $1.60^{\dagger}$ & 1.00 & 1.00 \\ \hline $S4.2$ &
0.00  & 0.09 & 506 & 2.13 & 0.29 & 0.99 & 0.98 \\ $S4.3$ & 0.10 &
0.26 & 50  & 1.50 & 0.97 & 0.98 & 0.98 \\
\hline
\label{tab:main}
\end{tabular}
\end{center}
\end{table}

Table~\ref{tab:main} lists for each simulation the following
information: the relative mass loss for the end-products $\Delta M
= (M_0 - M)/M_0$, the relative conservation of the global quantity
$Q$, with $\Delta Q =|Q_0 - Q|/Q_0$ referred to $\nu = 1/2$ unless
otherwise noted, the concentration $C_{\rho} = \rho(0)/\rho(r_M)$
of the end-products in terms of the ratio of the central density
to the density at the half-mass radius, the global anisotropy
parameter $\kappa = 2K_r/K_T$, the anisotropy radius (defined by
the relation $\alpha(r_{\alpha})=1$) relative to the half-mass
radius $r_{\alpha}/r_M$, and the final shape parameters $\epsilon$
and $\eta$. All quantities are referred to the final system of
bound particles.

\subsection{General properties}\label{sec:main}

\begin{figure}
  \resizebox{124pt}{!}{\includegraphics{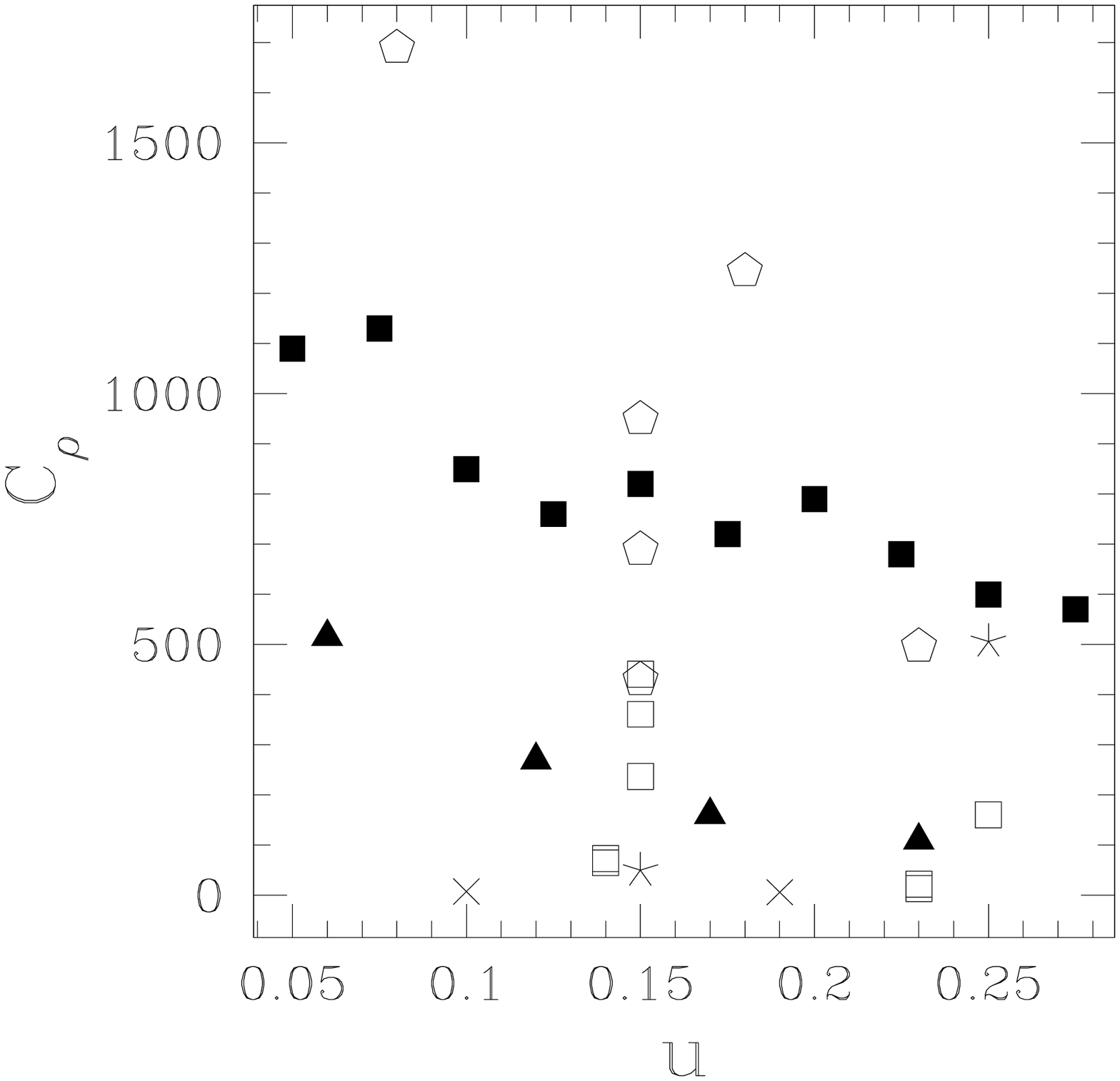}}
\resizebox{124pt}{!}{\includegraphics{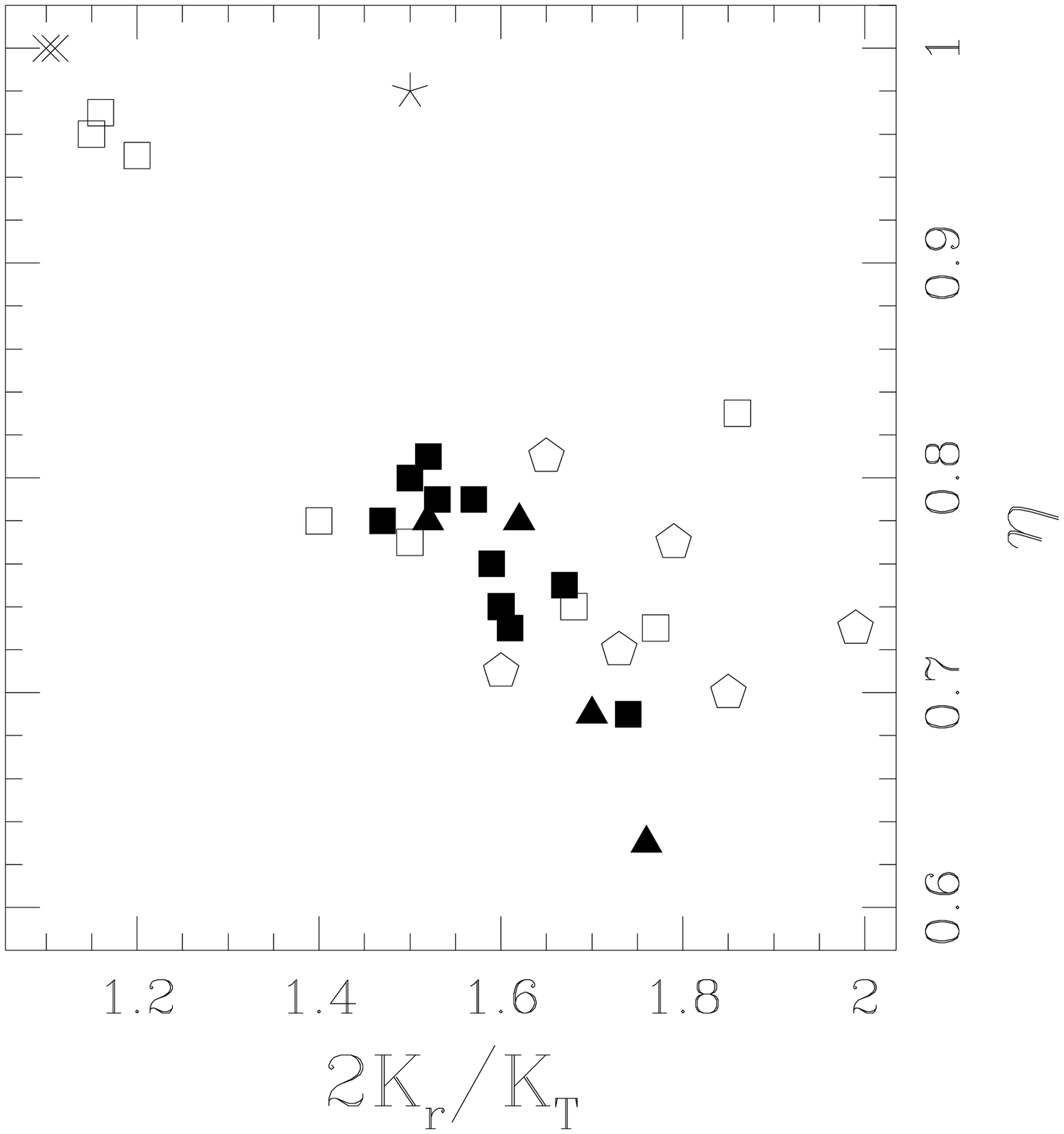}}
  \caption{Correlations between final concentration
$C_{\rho}$ and initial virial ratio $u$ (left), and between final
ellipticity $\eta$ and final global anisotropy $2K_r/K_T$ (right).
Symbols mark the various sets of simulations as follows: filled
squares for $C1$, filled triangles for $C2$, open pentagons for
$C3$, open squares for $C4$, crosses for $U$, and stars for $S$.}
  \label{fig:corr}
\end{figure}

From the results reported in Table~\ref{tab:main}, we may infer
some empirical trends. In particular, here we focus on: (1)
central concentration; (2) anisotropy content; (3) deviations from
spherical symmetry; (4) mass loss. Density and anisotropy profiles
will be discussed and compared with our theoretical models in
Section~\ref{sec:fit}.

We have run two series of simulations (type $C1$ and $C2$) for
which the initial particle positions and velocities are kept fixed
within each series, except for a constant scaling factor in the
velocities able to lead to different values of $u$ (from $0.05$ to
$0.275$ for $C1$ and from $0.06$ to $0.24$ for $C2$). This
procedure thus allows us to explore the role of the initial virial
ratio by keeping all other conditions strictly fixed.

The central concentration resulting from the collapse is expected to
correlate with $u$. \citet{lon91} proposed a simple criterion to set
an upper limit to the expected value of the central concentration by
imposing the conservation of the maximum density in phase space. They
argued that, for the collapse of an initially homogeneous system, the
central concentration, measured in terms of the ratio $r_M/r_{0.1}$
(of the half-mass radius to the radius of the sphere containing one
tenth of the total mass) should scale as $1/u$. Our $C1$ and $C2$
simulations follow qualitatively the proposed trend. However, since
relaxation is incomplete, it is natural to find that other factors, in
addition to the value of $u$, can contribute to determine the
properties of the final states. In fact, if we do not restrict our
attention to the $C1$ and $C2$ sequences only and consider instead the
entire set of simulations, we see that the correlation between $u$ and
$C_{\rho}$ becomes weaker (see Fig.~\ref{fig:corr}).

Differently from $C1$ and $C2$, the sets of $C3$ and $C4$
simulations, starting from different spatial configurations
(different number and size of clumps, different seed in the random
number generator), allow us to study other possible correlations,
in particular those between initial and final concentration and
between final concentration and initial deviations from spherical
symmetry; the latter correlation was noted by \citet{boi02},
starting from homogeneous spheroids. Again, if we include the
entire set of simulations, the correlations that we find are, in
general, relatively weak.

The final global anisotropy of the simulations (see the quantity
$\kappa$ in Table~\ref{tab:main}) is also weakly correlated with
$u$, with larger values of $u$ preferentially associated with
lower levels of radial anisotropy. The series $C1$ has a
systematic, but curiously non-monotonic trend, while $C3$ and $C4$
show that other factors, in addition to $u$, are important.

As to the shapes of the products of collisionless collapse, we
note a relatively strong correlation (see Fig.~\ref{fig:corr})
between the final shape (as measured by $\eta$) and the final
level of global anisotropy (as measured by $2K_r/K_T$). This is
likely to be related to the action of the radial orbit instability
during collapse. In particular, for the $C2$ series lower values
of $u$ lead to more anisotropic and more flattened end products;
the effect in the $C1$ series is less pronounced. Of course, the
issue of the final shapes produced by collapse has been addressed
by several investigations in the past, especially with the hope to
establish whether related dynamical mechanisms can account for the
observed morphologies of elliptical galaxies (for simulations in
the cosmological context, see \citealt{war92}; see also
\citealt{udr93}).

Initial conditions with a small number of clumps, as considered in
our paper, often show significant deviations from spherical
symmetry (from Table~\ref{tab:ic} we see that $\eta_0$ can be as
low as $0.7$). Curiously, the final value of $\eta$ may even
slightly exceed the value of $\eta_0$, thus showing that collapse
may sometimes push the system toward spherical symmetry, not
necessarily away from it.

Collisionless collapse can produce significant amounts of unbound
particles and consequently give rise to mass loss. This effect is
particularly severe in the cases where the collapse originates
from a homogeneous sphere \citep[see also ][]{lon91}; here the
system may lose up to one third of the mass (see run $U6.1$).
Clumpiness appears to have a stabilizing effect with respect to
mass loss; in fact, the mass lost is less than $7\%$ even for run
$C1.1$ characterized by $u=0.06$. On the other hand, symmetrized
clumpy initial states, of type $S$, are also found to evolve with
limited mass loss. [Since the nature of the gravitational forces
is mainly radial for both the collapsing homogeneous spheres ($U$
simulations) and symmetrized clumpy configurations ($S$), the
different amounts of mass loss might be related to the different
radial density distributions for the two types of run. In fact,
the effect of superimposing several clumps of particles creates a
density profile decreasing approximately linearly with radius.]

\subsection{The role of the radial orbit instability}

Spherical stellar systems with an excess of radial orbits ($2K_r/K_T >
1.7 \pm 0.25$) are expected to be unstable and to evolve rapidly, on
the dynamical time-scale, into ellipsoids; the precise value for the
onset of the radial-orbit instability depends on the detailed
structure of the system considered (\citealt{pol81}; see
\citealt{pal93}, and references therein).  The radial orbit
instability is thought to act efficiently during collisionless
collapse and is then argued to be the leading mechanism that makes
cold and spherical initial configurations evolve into generally
triaxial configurations \citep{agu90,pol92}.  The instability may also
be responsible for a reduction of the value of the central
concentration reached during collapse \citep{mer85b}; in fact, the
evolution of concentrated anisotropic systems into ellipsoids is
accompanied by a drastic softening of the density distribution
\citep{sti91}. As is the case for many other unstable systems,
evolution tends to remove the source of instability and thus, in our
case, to decrease the initial excess of radial orbits. Therefore, the
threshold of instability should provide an upper limit to the global
anisotropy of objects produced by collisionless collapse.

Our simulations largely confirm the general validity of this picture
and the general applicability of the \citet{pol81} criterion. In
particular, simulations $C2.3$ and $C2.4$ are characterized by a value
of $\kappa>1.7$ and lead to more flattened configurations than $C2.1$
and $C2.2$. Also the drop in the central concentration in simulation
$C1.10$ with respect to $C1.9$ might be related to the action of the
radial-orbit instability. Most of the end states are characterized by
relatively high anisotropy (generally $\kappa > 1.5$ and values around
$1.7$ are not infrequent) and thus it seems that evolution tends to
prefer a state very close to the stability boundary (as studied for
the $\f$ family of models in Paper I by means of an extensive set of
simulations). [An interesting finding is that symmetrized initial
conditions, although artificial, can lead to spherical final states
still able to sustain a large number of radial orbits ($\kappa \approx
2.1$ for simulation $S4.2$)].


\subsection{Angular momentum mixing}\label{sec:jscatter}

Simulations with homogeneous initial conditions generate
quasi-equilibrium final configurations that not only suffer from
significant mass loss, but also exhibit unusual features in their
anisotropy profiles (see Sect.~\ref{sec:ani} and Fig.~\ref{fig:uni2}).

\begin{figure}
  \resizebox{124pt}{!}{\includegraphics{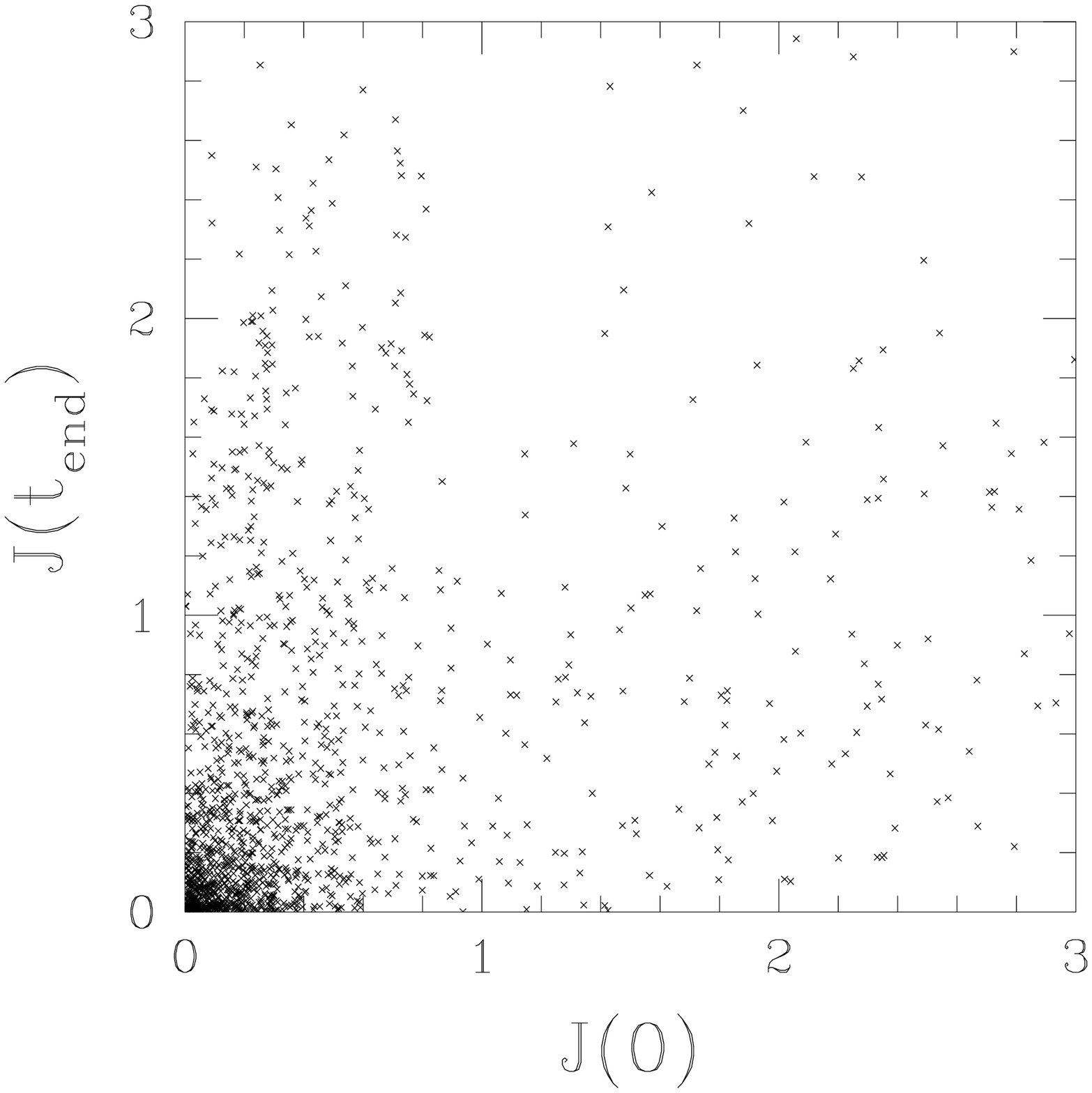}}
  \resizebox{124pt}{!}{\includegraphics{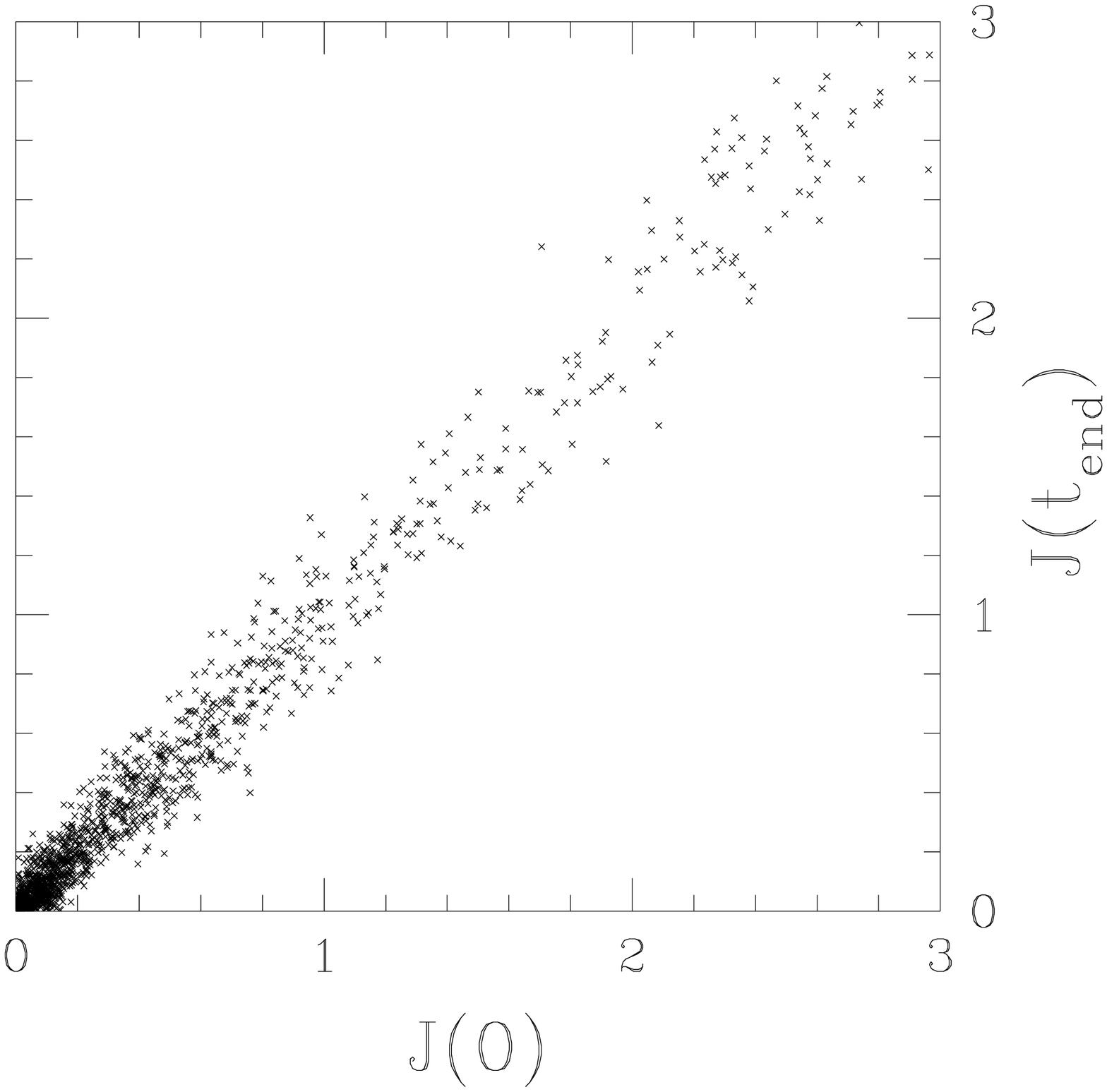}}
  \caption{Scatter plot (final vs. initial values) for the
  single-particle specific angular momentum. Comparison between
  a clumpy simulation (run $C4.2$; left panel) and its
  symmetrized version (run $S4.2$; right panel). Units for $J$ are
  $pc^2/yr$, see Sect.~\ref{sec:units}.}
  \label{fig:jscatter}
\end{figure}

If the degree of symmetry in the initial conditions is excessive,
little room is left for relaxation in the $(E,J^2)$ phase space even
if the process itself may be violent and lead to mass shedding. This
is confirmed by the fact that little or no mixing is observed in the
single-particle angular momentum distribution for homogeneous
simulations, as reported in Fig.~\ref{fig:jscatter} \citep[see
also][]{may84}. In fact, if the system evolves remaining close to
spherical symmetry, the conservation of single particle angular
momentum imposes severe constraints on the dynamical properties of the
end-state of the collapse. On the other hand, a certain degree of
clumpiness, even if limited to either position or velocity space,
leads to angular momentum mixing. This is confirmed by two test
simulations, $CV5.1$ and $CP5.2^*$, where mixing indeed turns out to
be quite efficient and leads to $J$ relaxation much like in the
left panel of Fig.~\ref{fig:jscatter} (see also Appendix).

Clumps thus help the system reach a ``universal'' final state from a
variety of initial conditions, which can explain the similarity of the
density profiles observed in the final products of collapse
simulations (see Sect.~\ref{sec:fit}).

\subsection{Dependence on the degree of clumpiness} \label{sec:clumps}

A few simulations with a large number of clumps ($400$ in $C4.5$
and $80$ in $C4.3$) and a spatial filling factor above unity
confirm that, in the limit of large $N_C$, the evolution of the
system approaches that of collapse simulations based on
homogeneous conditions, with end-states characterized by a flat
core and a low anisotropy content. A number of clumps of order
$10$ to $20$ thus seems to be optimal for an efficient violent
relaxation.

Even when limited to either position or velocity space, clumpiness
can be important and still lead to end-states with general
properties similar to those of the standard clumpy simulations
considered in this paper (see $CV5.1$ and $CP5.2$ entries in
Tables~\ref{tab:ic}-\ref{tab:main} and Sect.~\ref{sec:fit_cl}).

\begin{figure}
\resizebox{\hsize}{!}{\includegraphics{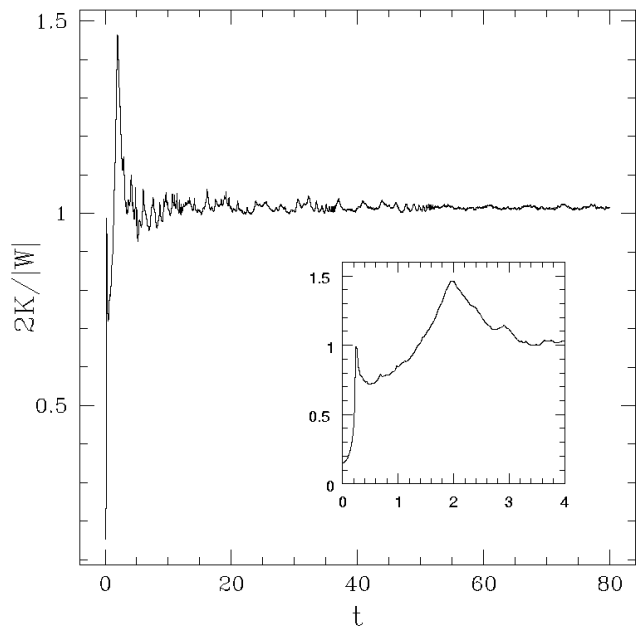}}
\caption{Evolution of the virial ratio during run $C4.4$,
characterized by the presence of many small cold clumps. The
insert box zooms in on the evolution at the beginning of the
simulation, when a first collapse occurs, followed by an expansion
of the clumps while collapsing toward the center of mass of the
system. The virial ratio in the final stages of the simulation is
slightly above unity because of mass loss.}
  \label{fig:smallcl}
\end{figure}

We also studied the dependence of the results of collisionless
collapse on the spatial filling factor of the clumps. To do this,
we took advantage of the ability of \emph{GyrFalcON} to deal with
systems with different scales and ran a simulation ($C4.4$)
initialized with $80$ {\it small} cold clumps (i.e. with a radius
$R_{C}=2.8~kpc$ distributed in a sphere of radius $40~kpc$). For
this simulation, evolution basically occurs in two stages, with a
first collapse in which strongly bound structures are formed in a
very short time, followed by subsequent merging (see
Fig.~\ref{fig:smallcl}). Interestingly, the outcome of this
simulation is highly isotropic ($\alpha \approx 0$ out to the
half-mass radius) and very concentrated. We will see
(Sect.~\ref{sec:fit_cl}) that, even in this case, the density
profile remains very well represented by $\f$ models (and by the
$R^{1/4}$ law). For $C4.4$, after several tens of dynamical times,
there remain traces (remnants) of the more strongly bound clumps,
orbiting within the smooth system.

\section{Conservation of $Q$} \label{sec:Q}

We recall that the quantity $Q$ (for a discrete system of $N$
particles $Q=\sum_{i=1,N} (J_i/|E_i|^{3/4})^{\nu}$; see also
Sect.~\ref{sec:units}) has been introduced for the description of
conditions in which partial violent relaxation occurs, where it is
argued that information about the initial state is basically lost,
except for an approximate conservation of a third quantity (in
addition to total energy and total number of particles). Therefore, it
would be wrong to invert the argument and imagine that, by itself, the
conservation of a quantity such as $Q$ is equivalent to the picture of
incomplete violent relaxation. In particular, we note that, judging
from our set of simulations, $Q$ is well conserved for homogeneous
initial conditions, both in the velocity and position space. However,
this is less relevant to our goals, since homogeneous conditions do
not allow for mixing and violent relaxation at the level of angular
momentum space to operate properly. Therefore, it is not surprising to
find that the end-products of simulations with homogeneous initial
conditions tend to be less well represented by the $\f$ models, in
spite of their relatively good conservation of $Q$.

In this Section we will show that the issues involved in the
conjectured conservation of $Q$ and the indications obtained from
our simulations are complex. Therefore, it would be pointless
to continue further in this direction, looking for a better
definition of what might be defined as ``acceptable degree of
conservation" or searching for other quantities that might be
conserved better than $Q$. Instead, to make a decisive test about
the merits of our approach, we should take the models that have
been constructed (by means of the {\it Ansatz} of the
$Q$-conservation) and compare them in detail with the results of
collisionless collapse obtained from our simulations. Such test
will be addressed in the following Sect.~\ref{sec:fit}.

\subsection{The ``observed'' conservation}

The value of $Q$, computed with $\nu = 1/2$, is approximately
conserved for a wide range of initial configurations. By
approximate conservation we mean that $\Delta Q \leq 0.5$,
although in some cases we have conservation as good as $\Delta Q
\approx 0.01$. As a general rule, $Q$ is better conserved if the
initial virial ratio is not too low. 

A curious property is that all clumpy simulations appear to lead to
the {\it same} value of $Q$, with a scatter on the order of $10 \%$
(see Table~\ref{tab:fit}). [I.e. the scatter is \emph{less} than the
mean deviation from exact conservation, around $20-30 \%$.] This
result can be interpreted, at the level of the simulations, by
considering that, independently of the specific details of the initial
clumpy conditions, the large scale structure of the end products of
the simulations is very similar, with respect both to physical scales
(constrained by the conservation of mass and energy in the collapse) as
well as dimensionless dynamical properties at large radii (see
Sect.~7). In addition, the fact that the values of $(M,E_{tot},Q)$
realized at the end of the simulations are approximately constant is
consistent with the fact that the best-fit models do not exhibit wide
variations in the values of $\Psi$ and $\nu$ (cf. Table~3 and the
discussion of parameter space given by Bertin \& Trenti 2003,
Sect.~3).

Strict conservation is not meaningful, for a number of reasons.
Indeed, during collisionless collapse even the total number of
particles $N$ and the total energy $E_{tot}$ are not conserved, if we
refer these quantities to the final set of {\it bound particles}; it
was noted \citep{sti87} that the non-conservation of $Q$ actually
correlates with the non-conservation of $N$ and $E_{tot}$. A simple
argument also warns us that the conservation of $Q$ should not be
meant to apply to all conditions. The reason is that, if we refer to
the proposed definition, $Q$ cannot be conserved in the limit of an
infinitely cold collapse. In fact, for an infinitely cold collapse
(i.e. for $ u \rightarrow 0$, with the stars kept at fixed initial
positions), at the beginning of the simulation we would have $Q
\rightarrow 0$ (because the single-particle angular momenta vanish, in
the limit of vanishing initial velocities, while the single-particle
binding energies remain at a finite value). On the other hand, at the
end of the simulation, the formation of a quasi-isotropic core with
finite kinetic energy content requires that the final value of $Q$ be
finite.

Furthermore, the quantity $Q$ is referred to an ideal case
characterized by spherical symmetry, while, as noted earlier, both
the initial and the final configurations in our simulations of
collisionless collapse can exhibit significant deviations from
spherical symmetry. To get an estimate of changes of $Q$
associated with deviations from spherical symmetry, we have
considered a $(1/2;3)$ $\f$ model, unstable against the radial
orbit instability (Paper I), and let it evolve; the final
quasi-equilibrium state is characterized by $\epsilon \approx \eta
\approx 0.73$ and is associated with a change $\Delta Q = 0.12$.
Similar changes are observed by stretching artificially an $\f$
model to a non-spherical geometry, with $\epsilon = 1$ and $\eta
\approx 0.7$. But these changes are given for comparison only,
since they are not related to conditions in which violent
relaxation takes place.

\subsection{General polynomial dependence}

Although aware of the fact that we should not really look for
quantities conserved exactly during collisionless collapse, we
decided to test the paradigm of $Q$ conservation further by
considering the more general class of $Q$ functionals, defined,
for a system of $N$ points, as:

\be \label{eq:Qgen}
\tilde{Q}=\sum_{i=1}^{N}\frac{J_i^{\nu_2}}{|E_i|^{\frac{3}{4}\nu_1}},
\ee

\noindent where $\nu_1$ and $\nu_2$ are free parameters. We
explored the parameter space $-1 \leq \nu_1 \leq 1$ and $-1 \leq
\nu_2 \leq 1$. The functional $Q$ used to construct the $\f$
models corresponds to the condition $\nu= \nu_1 = \nu_2 > 0$,
which guarantees the desired asymptotic behavior for the
associated density $\rho \sim r^{-4}$ at large radii.

\begin{figure}
  \resizebox{\hsize}{!}{\includegraphics{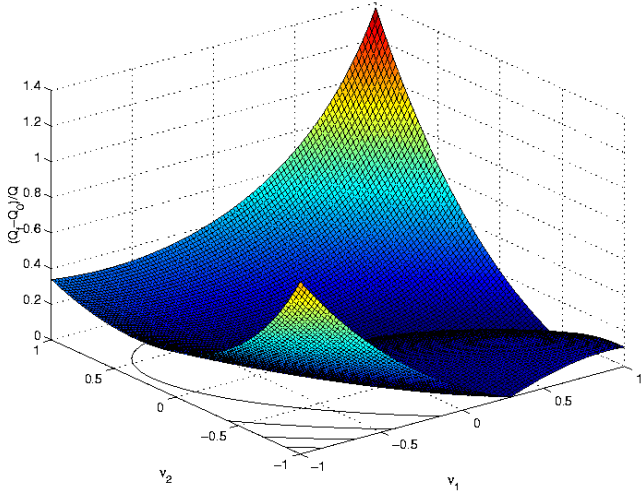}}
\caption{Conservation of the general functional $\tilde{Q}$
(see Eq.~\ref{eq:Qgen}) in a typical simulation ($C4.1$).}
  \label{fig:Qgen}
\end{figure}

We studied the change in the value of this functional computed at the
beginning and at the end of a typical simulation ($10^5$ particles in
$10$ cold clumps, run $C4.1$; see Fig.~\ref{fig:Qgen}). If we focus on
the $\nu_1 = \nu_2 = \nu$ condition, the best conservation would be
attained for low values of $\nu$.

\section{Fit with the $\f$ models}\label{sec:fit}

We first fit the density and the pressure anisotropy profiles,
$\rho (r)$ and $\alpha (r)$, of the end-products of our
simulations by means of the $f^{(\nu)}$ family of models. The
phase space properties of the best-fit model thus identified are
then compared with those of the end-products of the simulations.

Smooth, angle-averaged simulation profiles are obtained by binning
the particles in spherical shells and averaging over time, based
on a total of $20$ snapshots taken from $t = 64$ to $t = 80$, at
an epoch when the system has already settled down in a
quasi-equilibrium configuration. For the $\f$ models, the
parameter space explored is that of an equally spaced grid in
$(\nu,\Psi)$, with a subdivision of $1/8$ in $\nu$, from $3/8$ to
$1$, and of $0.2$ in $\Psi$, from $0.2$ to $14.0$ (corresponding
to the grid of models studied in Paper I). The mass and
the half-mass radius of the models are fixed by the scales set by
the simulations.

A minimum-$\chi^2$ analysis is then performed, with error bars
estimated from the variance in the time average process used to
obtain the smooth simulation profiles. A critical step in this
fitting procedure is the choice of the relative weights for the
density and the pressure anisotropy profiles. We adopted equal
weights for the two terms, checking a posteriori that their
contributions to $\chi^2$ are of the same order of magnitude.

\begin{table}[!h]
\caption{Best fit $\f$ models for the set of high resolution runs
(series $C2$ and $C3$). The various columns give: run identifier,
model identifier, mean value of the absolute relative deviations
from the density of the simulations, mean value of the absolute
deviations in the pressure anisotropy profile, mean value of the
absolute relative deviations in the energy distribution, and final
value of $Q$.}
\begin{center}
\begin{tabular}{cccccc}
\hline \hline &  $\f$ & $\langle |\Delta \rho / \rho | \rangle $ &
$\langle |\Delta \alpha| \rangle $ & $\langle |\Delta E / E| \rangle$
& $Q$\\ \hline


$C2.1$ & (1/2;4.8) & 0.11 & 0.07 & 0.23 & 1.24 \\    
$C2.2$ & (1/2;4.8) & 0.11 & 0.06 & $0.22$ & 1.33 \\    
$C2.3$ & (5/8;5.0) & 0.12 & 0.06 & 0.21 & 1.35 \\
$C2.4$ & (7/8,5.6) & 0.14 & 0.08 & $0.23$ & 1.33 \\

$C3.1$ & (3/8;5.6) & 0.10 & 0.22 & $0.18$ & 1.33 \\
$C3.2$ & (3/8;5.4) & 0.11 & 0.19 & $0.22$ & 1.26 \\  
$C3.3$ & (1/2;5.2) & 0.17 & 0.16 & $0.20$ & 1.64 \\ $C3.4$ &
(5/8;5.4) & 0.12 & 0.05 & 0.18 & 1.40 \\ $C3.5$ & (1/2;6.2) & 0.09
& 0.20 & 0.15 & 1.35 \\
$C3.6$ & (3/8;5.2) & 0.13 & 0.05 & $0.20$ & 1.35 \\    

\hline
\label{tab:fit}
\end{tabular}
\end{center}
\end{table}

\subsection{Density profiles}

Since the half-mass radius $r_M$ and the total mass $M$ are kept fixed
in the fitting procedure, we are left with two degrees of freedom
(i.e., the dimensionless parameters $\nu$ and $\Psi$). In practice,
given the general behavior of the density profile of the $\f$ models
(see Fig.~3 in Paper I), at large radii the freedom in the fit
is limited. Therefore, the excellent match at large radii to the
density profile of the end-products of the simulations demonstrates
that the $\f$ family has been constructed on solid physical
grounds. Different values of $(\nu,\Psi)$ correspond to different
shapes of the inner potential well and of the anisotropy profile. As
exemplified by Figs.~\ref{fig:C3.5}-\ref{fig:C3.4}, the density of the
final systems produced by the high resolution set of simulations ($C2$
and $C3$) is well represented by the best-fit $f^{(\nu)}$ profile over
the entire radial range, from $0.1$ to $10$ half mass radii. The fit
is satisfactory not only in the outer parts, where the density falls
by {\it nine orders of magnitude} with respect to the central regions,
but also in the inner regions. The mean absolute relative deviation
between simulations and models ($\langle |\Delta \rho / \rho| \rangle
=(1/N_g)\sum_{i=1}^{N_g} |\rho_{sim}(r_i) -\rho_{model}(r_i)|/
\rho_{sim}(r_i)$), computed over this extended radial range, is
usually around $10 \%$ (see Table~\ref{tab:fit}); here $N_g$
represents the number of radial grid points.

With a similar procedure, we have studied the end-products of
simulations characterized by different numbers of particles and
clumps ($C1$ and $C4$). No significant changes in the quality of
the fits are found if we focus on simulations characterized by
clumpy initial conditions (with the possible exception of those
run with $N_C \ge 80$) .

\subsection{Projected density profiles}

The end-products of collisionless collapse are known to be
characterized by projected density profiles generally well fitted
by the $R^{1/4}$ law \citep{dev48}, provided that the collapse
factor is large (i.e., that the initial virial ratio $u$ is small;
see \citealt{van82,lon91}). With our set of simulations we confirm
this result and we extend it by means of the $\f$ models.

The successful comparison between models and simulations is
interesting because, depending on the value of $u$, some simulations
lead to configurations that exhibit deviations from the $R^{1/4}$
law. In these cases, the density profile projected along the
line-of-sight is characterized by an $R^{1/n}$ behavior with $n \neq
4$.  For example the $C2.4$ simulation, which starts with a low
collapse factor, has a best fit index $n \approx 3$, while the the
simulation $C3.1$, which has a large collapse factor, is best
represented by a profile with $n \approx 5$. Yet, these systems turn
out to be all well fitted by the $\f$ models. Therefore, the family of
models that we have identified might also be useful to describe
systematic structural changes in galaxies, in the framework of the
proposed weak homology of elliptical galaxies \citep{ber02}.

\subsection{Pressure anisotropy profiles} \label{sec:ani}

\begin{figure}
  \resizebox{\hsize}{!}{\includegraphics{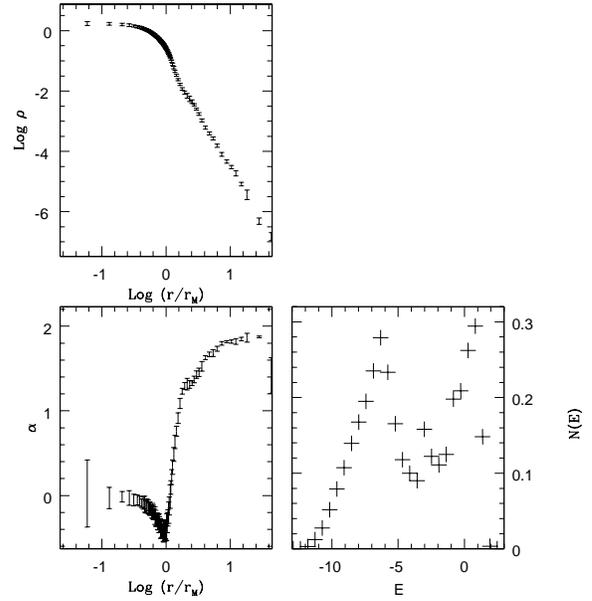}} \caption{ Density
and anisotropy profiles (left frames) and energy density
distribution (right frame) for simulation $U6.2$, starting from a
homogeneous sphere. Note that in the vicinity of the half-mass
radius the pressure anisotropy is {\it tangentially} biased.}
  \label{fig:uni2}
\end{figure}

In our simulations the pressure anisotropy profiles follow the general
trend expected for the process of collisionless collapse.  In
particular, the final configurations are characterized by an isotropic
core, with $\alpha \approx 0$, while the outer regions have a strongly
radially biased anisotropy (up to $\alpha=2$). The transition region
($\alpha \approx 1$) is located around the half-mass radius (see
column $r_{\alpha}/r_M$ in Table~\ref{tab:main}). Higher values of
$2K_r/K_T$ are associated with lower values of $r_{\alpha}/r_M$. For
clumpy initial conditions (with the possible exception of those run
with $N_C \ge 80$), the anisotropy profile $\alpha(r)$ is a monotonic
increasing function of the radius. A curious feature is found for the
results of collapse of uniform spheres (runs $U$). Here (see
Fig.~\ref{fig:uni2}) the core is basically isotropic, with the region
around the half-mass radius exhibiting an excess of {\it tangential}
orbits (up to $\alpha \approx -0.4$). In the outer parts, but with a
very sharp transition, the pressure profile becomes radially
biased. In correspondence to the dip in $\alpha$, where $\alpha<0$, we
note a clear feature in the density profile (see
Fig.~\ref{fig:uni2}). Uniform spheres initialized with a very small
particle number ($N < 10^4$) do not show this behavior; for them the
pressure anisotropy rises quite regularly, although the profile is
significantly affected by Poisson noise.

In conclusion, for all the clumpy $C$ runs (again, we should
mention, with the possible exception of those runs with $N_C \ge
80$), the anisotropy profile is represented extremely well by our
models, with a mean absolute error ($\langle |\Delta \alpha|
\rangle = (1/N_g)\sum_{i=1}^{N_g} |\alpha_{sim}(r_i)
-\alpha_{model}(r_i)|$) typically around $0.1$ but often as low as
$0.05$ (see Table~\ref{tab:fit}).

\begin{figure}
  \resizebox{\hsize}{!}{\includegraphics{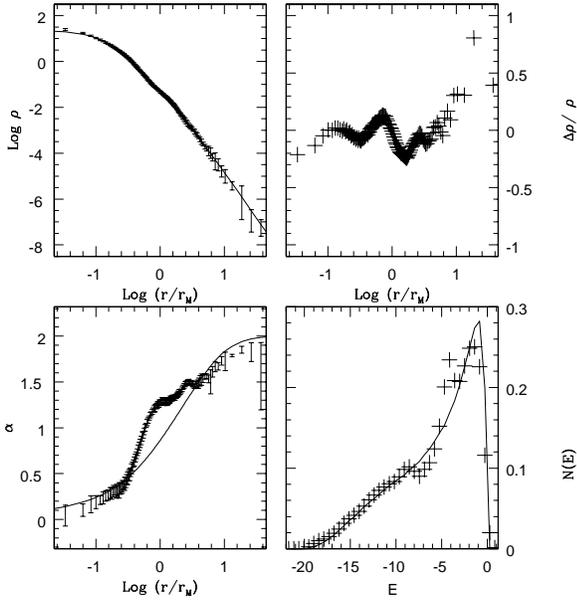}} \caption{
Comparison between the $C3.5$ simulation and the best-fit
$f^{(\nu)}$ model ($1/2;6.2$). The top left panel represents the
density as measured from the simulation (error bars) and the
best-fit profile (line). The top right panel gives the residuals
from the fit. At the bottom left, the anisotropy profile of the
simulation (error bars) is compared with the best-fit profile
(line); the bottom right frame illustrates the energy density
distribution $N(E)$. The density $\rho$ and the single-particle
energy $E$ are given in code units.}
  \label{fig:C3.5}
\end{figure}

\begin{figure}
  \resizebox{\hsize}{!}{\includegraphics{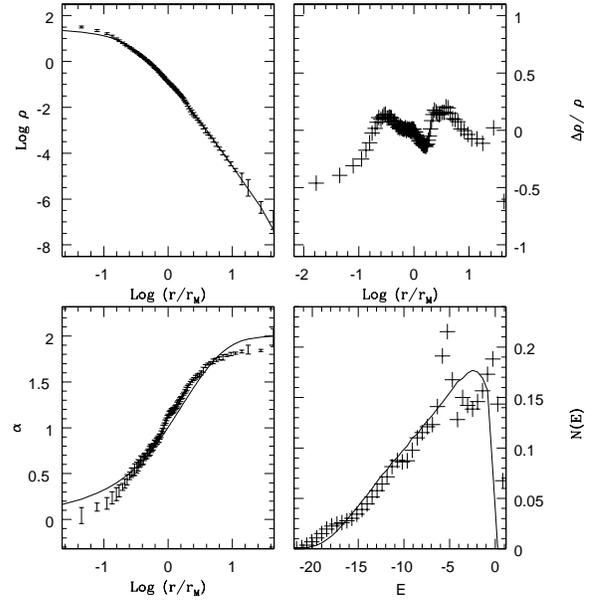}}
\caption{Comparison between the $C2.1$ simulation and the best-fit
$f^{(\nu)}$ model ($1/2;4.8$), shown as in Fig.~\ref{fig:C3.5}.}
  \label{fig:C2.1}
\end{figure}

\begin{figure}
  \resizebox{\hsize}{!}{\includegraphics{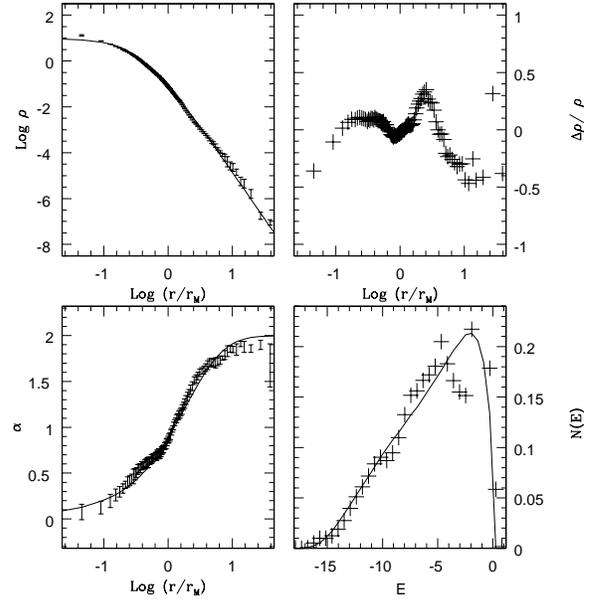}} \caption{
Comparison between the $C3.4$ simulation and the best-fit
$f^{(\nu)}$ model ($5/8;5.4$), shown as in Fig.~\ref{fig:C3.5}.}
  \label{fig:C3.4}
\end{figure}

To some extent, the final anisotropy profiles for clumpy initial
conditions are found to be sensitive to the detailed choice of
initialization. In other words, runs starting from initial conditions
with the same parameters, but with a different seed in the random
number generator, give rise to slightly different profiles. In any
case, the agreement between the simulation and the model profiles
remains very good (see Figs.~\ref{fig:C3.5}-\ref{fig:C3.4}).



\subsection{Comparison at the level of phase space}

At the level of phase space, we have performed two types of
comparison, one involving the energy density distribution $N(E)$
and the other based on $N(E,J^2)$. The chosen normalization
factors are such that: \be\label{eq:ne} M = \int N(E) dE = \int
N(E,J^2) dE dJ^2. \ee \noindent

The energy distributions $N(E)$ that we find (see
Fig.~\ref{fig:C3.5}-\ref{fig:C3.4}), qualitatively similar to
those obtained in earlier investigations (see Fig.~2 in
\citealt{van82} and Fig.~10 in \citealt{udr93}), are characterized
by an approximate exponential behavior at low energies ($N(E)
\propto \exp{(-aE)}$) with a rapid cut-off near the origin, which
is argued to go as $|E|^{5/2}$ because the potential is Keplerian
in the outer parts (\citealt{udr93}; see also the discussion by
\citealt{jaf87} and by \citealt{ber89}). The final states of the
simulations also show the presence of particles with positive
energy, escaped from the system.

In Fig.~\ref{fig:C3.5} we plot the final energy density distribution
for the simulation run C3.5 with respect to the predictions of the
best-fit model identified from the study of the density and pressure
anisotropy distributions. Similar plots are given in the following
figures for other simulations. The agreement is very good ($\langle
|\Delta E| \rangle \approx 0.2$, see Table \ref{tab:fit}), especially
for the strongly bound particles. In particular, this means that we
are correctly describing the innermost part of the system. The energy
distribution for less bound particles (i.e. those associated mostly
with the outer parts of the system) is less regular and sometimes
presents a double peak (e.g., see Fig.~\ref{fig:C2.1}), which
obviously cannot be matched in detail by our models. This is an
interesting example of the way some memory of the initial state can be
preserved (the extra-peak is indeed related to the initial
distribution of binding energies) and a direct sign of the
incompleteness of violent relaxation.

\begin{figure}
  \resizebox{\hsize}{!}{\includegraphics{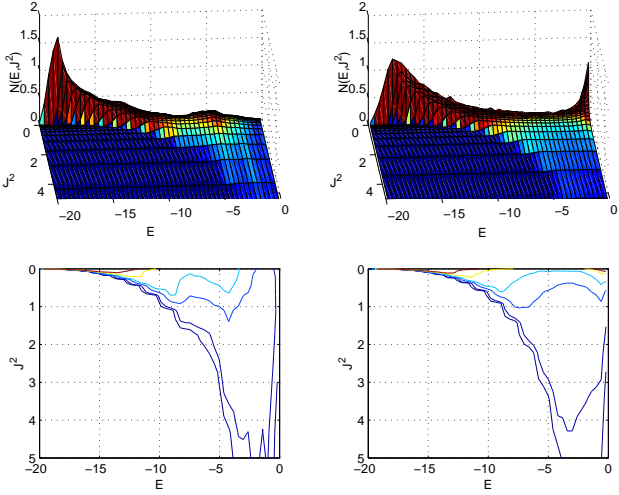}}
\caption{Final phase space density $N(E,J^2)$ (left column) for
the simulation $C3.5$, compared with that of the best fitting
$(1/2;6.2)$ $f^{(\nu)}$ model (right column).}
  \label{fig:NEJ2_c3.5}
\end{figure}

\begin{figure}
  \resizebox{\hsize}{!}{\includegraphics{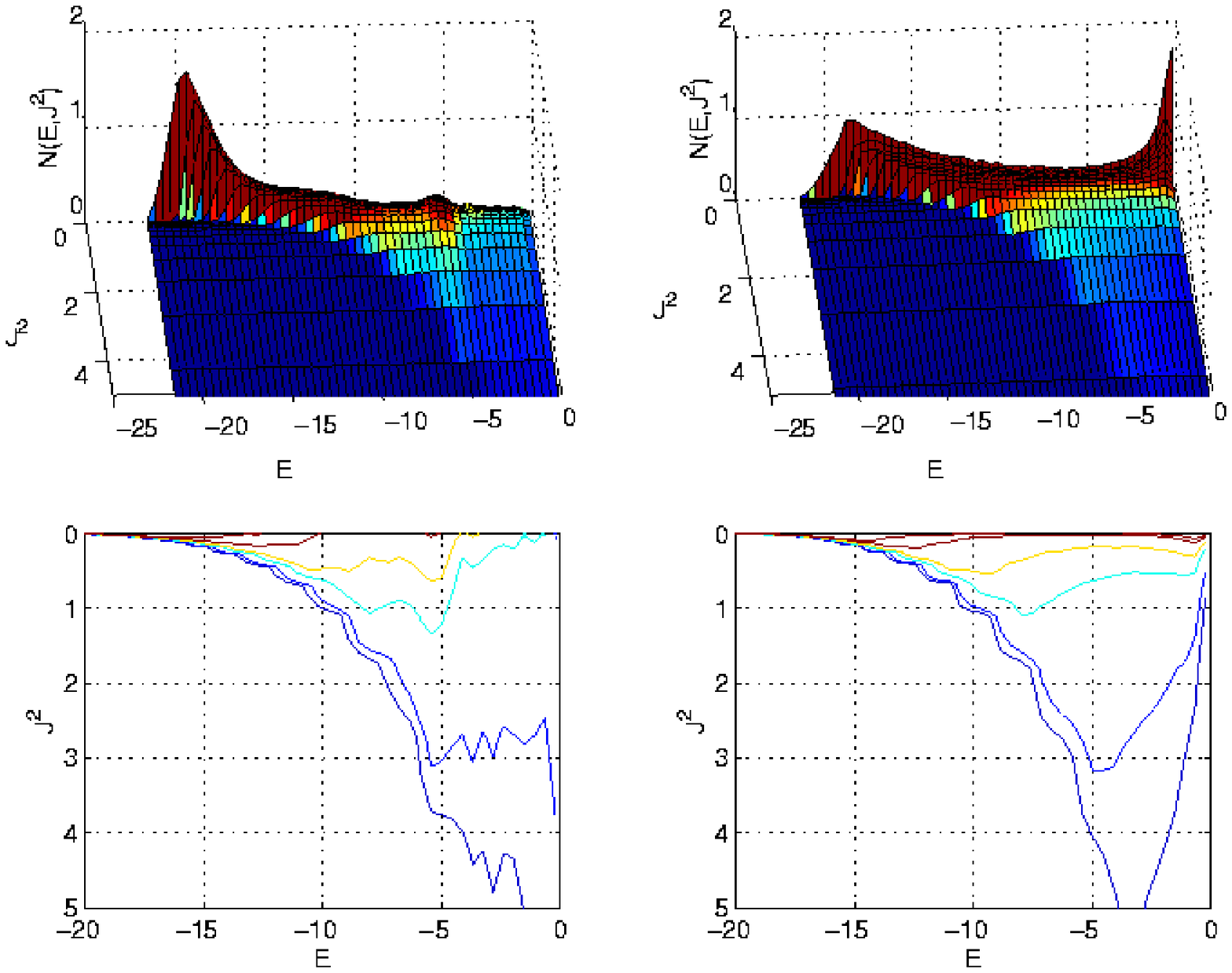}}
\caption{Final phase space density $N(E,J^2)$ (left column) for
the simulation $C2.1$, compared with that of the best fitting
$(1/2;4.8)$ $f^{(\nu)}$ model (right column).}
  \label{fig:NEJ2_c2.1}
\end{figure}

\begin{figure}
  \resizebox{\hsize}{!}{\includegraphics{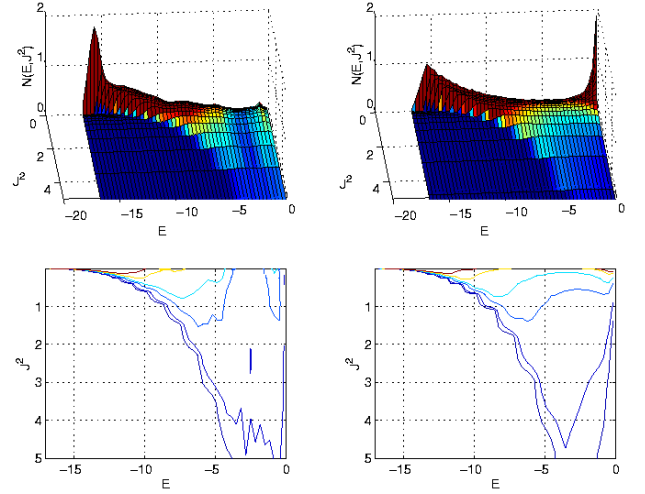}}
\caption{Final phase space density $N(E,J^2)$ (left column) for
the simulation $C3.4$, compared with that of the best fitting
$(5/8;5.4)$ $f^{(\nu)}$ model (right column).}
  \label{fig:NEJ2_c3.4}
\end{figure}

Finally, at the deeper level of $N(E,J^2)$, simulations and models
also agree rather well, as illustrated in the four panels of
Figs.~\ref{fig:NEJ2_c3.5}-\ref{fig:NEJ2_c3.4}. For the cases
shown, the distribution contour lines are in good agreement in the
range from $E_{min}$ to $E \approx - 4$; however, the theoretical
models show a peak located near the origin, not present in the
simulations, which is related to the Jacobian factor arising from
the transformation of the $\f$ distribution function from the
($\vec{x},\vec{w}$) to the ($E,J^2$) space.

\subsection{An additional test to characterize clumpy initial
  conditions} \label{sec:fit_cl}

As an additional test to characterize the detailed effects of
clumpiness, we studied the end-products of the $CV5.1$ and
$CP5.2^*$ simulations, by comparing them with the $\f$ models.

Although these two runs start from initial conditions rather different
from our standard choice (cf. $C1$-$C3$), being homogeneous either in
position ($CV5.1$) or in velocity ($CP5.2^*$) space, we note that they
can be fitted very well by our family of models: $(3/4;5.4)$ for
$CV5.1$ and $(1;6.2)$ for $CP5.2^*$ (with $\langle |\Delta \rho / \rho|
\rangle \approx 0.1$). The good match at the level of the anisotropy
profile $\alpha(r)$ and of the single-particle energy distribution
also confirms, as discussed in the Appendix, that the requirement of
clumpiness in phase space is a well posed characterization of the
initial conditions. The two runs have the following behavior with
respect to the $Q$-conservation: $\Delta Q = 0.02$ and final value $Q
= 1.40$ for $CV5.1$; $\Delta Q = 0.01$ and final value $Q =1.26$ for
$CP5.2^*$.

In passing, we note that the $C4.4^*$ simulation, characterized by
very small clumps, leads to a concentrated final density profile
that is well reproduced by the $(1;9.2)$ $\f$ model (with $\langle
|\Delta \rho / \rho| \rangle \approx 0.15$).


\subsection{Separate fits to density and anisotropy profiles by means of simple analytic functions}

\begin{figure}
\resizebox{124pt}{!}{\includegraphics{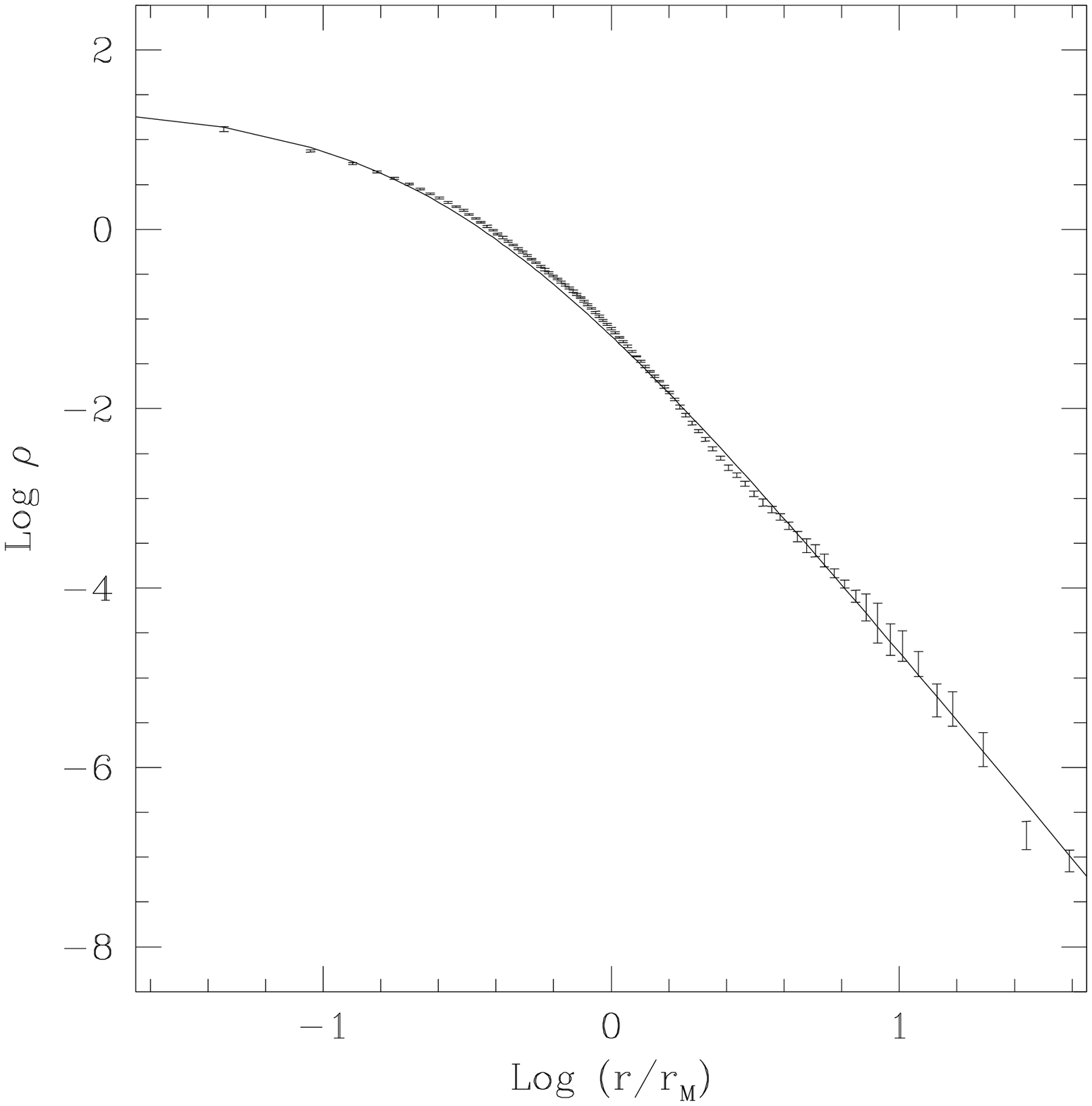}}
\resizebox{124pt}{!}{\includegraphics{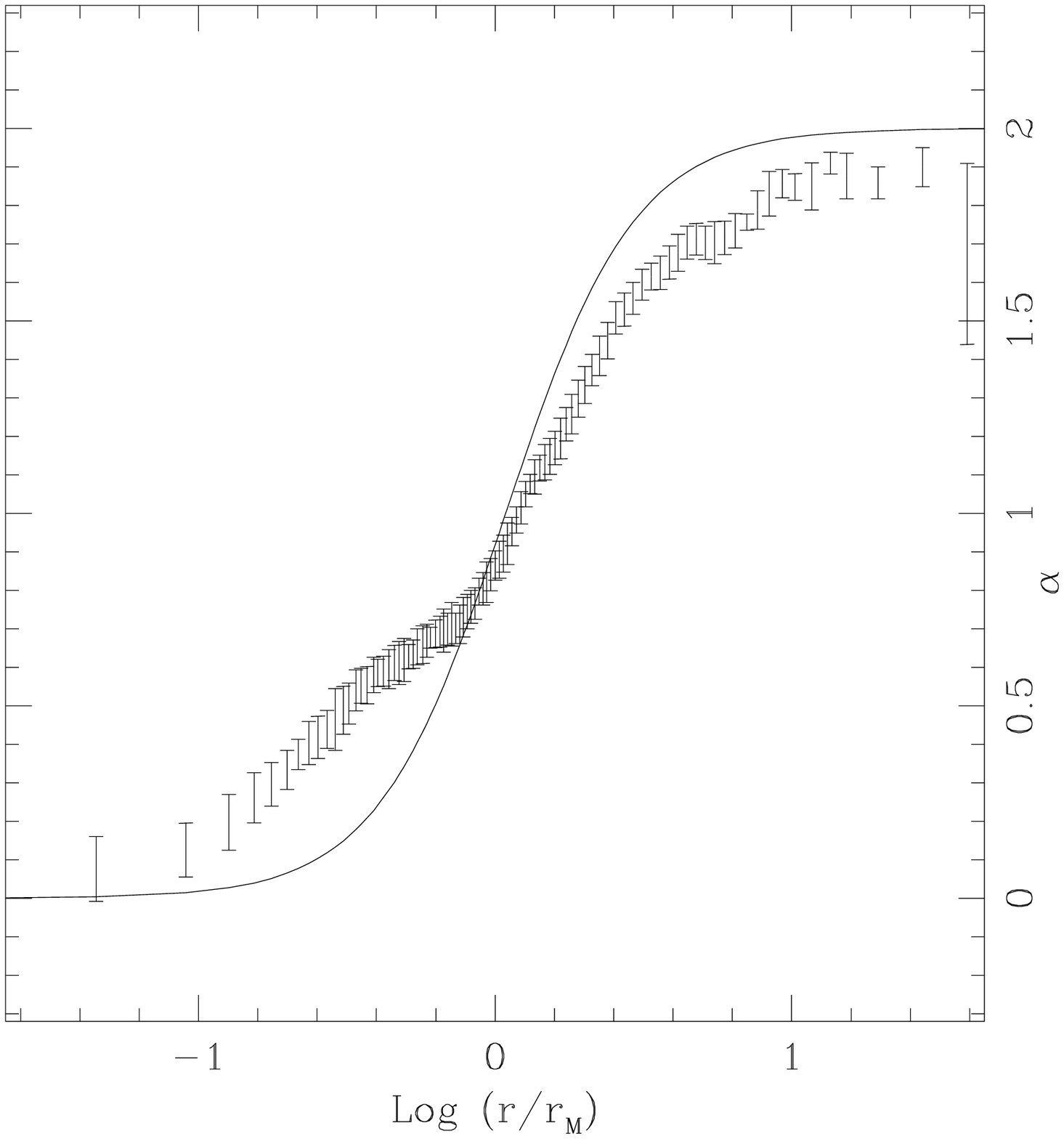}}
\caption{ Density profile (left), fitted using Eq.~(\ref{eq:deh_rho})
  with $\gamma=0.11$, and anisotropy profile (right), fitted with
  Eq.~(\ref{eq:OM}), for the simulation $C3.4$. Compare to the fit
  with the (5/8;5.4) $\f$ model in Fig.~\ref{fig:C3.4}.}
  \label{fig:OM}
\end{figure}

Simple analytic descriptions of density profiles and, separately, of
anisotropy profiles are often used in stellar dynamics, without a
specific physical scenario of galaxy formation. For the density
profile we may refer to: 
\be \label{eq:deh_rho} 
\rho(r) =
\frac{(3-\gamma)M}{4 \pi} \frac{r_0}{r^{\gamma}(r+r_0)^{4-\gamma}}, 
\ee
where $0 \leq \gamma < 3$ is a free parameter, and $M$ and $r_0$ are a mass
and length scale respectively \citep{deh93}. As discussed in Paper I,
it is no surprise to find that the case $\gamma = 2$ \citep{jaf83}
captures the general properties of the density profile obtained by the
simulations at the $20 \%$ level.  Curiously, when we fit the density
distribution of some simulations by means of Eq.~(\ref{eq:deh_rho}),
the best fitting index $\gamma$ is very low $\gamma \approx 0.1$ (see
Fig.~\ref{fig:OM}).

Similarly, for the anisotropy profile one might resort to the analytic
distribution \be \label{eq:OM} \alpha(r) = 2 \frac{r^2}{r^2+r^2_{\alpha}}, \ee with
$r_{\alpha}$ being a free scale \citep{mer85c}. As shown in
Fig.~\ref{fig:OM}, the typical shape of the anisotropy profile reached
at the end of the simulations is different.

\section{Discussion and conclusions} \label{sec:con}

In this paper we have concentrated on nearly spherical,
one-component stellar systems. As is well known, in spite of these
restrictions, the equations of stellar dynamics allow for almost
complete freedom in the construction of self-consistent dynamical
models, with the only requirement that they should be supported by
a positive definite (but otherwise arbitrary) function of $E$ and
$J$, as a distribution function in phase space. Therefore, the
full range of self-consistent one-component spherical stellar
dynamical models is enormous. Most likely, the majority of these
models have little to do with the systems that have been realized
in nature. The main idea at the basis of the present paper is to
combine clues from $N$-body simulations and from statistical
arguments so as to pinpoint, among the enormous variety of in
principle acceptable dynamical models, those few that, because of
their physical justification, have a chance of matching the
properties of interesting classes of numerical simulations and of
observed stellar systems.

Some interesting clues had been noted earlier. With the aim of
summarizing the main properties of incomplete violent relaxation
during collisionless collapse, it was discovered \citep{sti87}
that, by arguing that a third quantity $Q$ (in addition to total
energy and number of stars) should be included among the relevant
constraints in the extremization of the Boltzmann entropy, the
most probable and thus physically justified distribution function
$\f$ leads to models that are in {\it general qualitative}
correspondence with the products of collisionless collapse found
in numerical simulations and with the observed luminosity profiles
of bright elliptical galaxies.

In the present paper we have demonstrated that the $\f$ models are
able to match in  {\it surprising quantitative detail} the results
of our numerical simulations. At the same time, the $\f$ models
exhibit projected density profiles that are well represented by
the $R^{1/n}$ law (generally with $n \approx 4$; the residuals
from the fit are within $0.1$ magnitudes in a radial range from
$0.1$ to $10$ effective radii; see also Paper I).
Therefore, we have demonstrated that the $\f$ models, as well as
the end products of the collapse simulations, are relevant to the
description of the stellar distribution of elliptical galaxies.
Such correspondence is even more remarkable, if we recall that,
from the results established in the last decades, dark matter
should play a dominant role in the structure of galaxies, while
our approach neglects, so far, some important ingredients among
which the presence of a massive, possibly diffuse dark halo.

Independently of stellar dynamical modeling, our simulations have
shown that clumpy initial configurations allow for an efficient
re-distribution of the angular momenta of the individual particles
during collapse: such efficient phase space mixing is precisely
the main condition required for a successful application of the
statistical arguments that lead to the construction of the $\f$
family of distribution functions. In the past \citep[e.g.,
see][]{van82,may84,mer85b,lon91}, it has been noted that cold
collapses, within a wide range of initial density profiles,
generate quasi-equilibrium systems with approximate $R^{1/4}$
profiles. Here we confirm that the best match to approximate
$R^{1/4}$ profiles is obtained from initiallly clumpy
configurations. It thus appears that collapses starting from
artificially uniform and spherically symmetric initial conditions
retain too much memory of the initial conditions and are unable to
evolve into a universal density distribution. Therefore, it is
interesting to find that precisely those initial conditions that
look more plausible and realistic from the physical point of view
lead to end products able to match the stellar distribution of
observed systems in detail. We may then conclude that
collisionless collapse from clumpy initial conditions followed by
violent relaxation is indeed a formation mechanism relevant to
elliptical galaxies.

If we now take the point of view of stellar dynamical modeling and
examine the foundation of the $\f$ family of models, we note that
many collapse simulations show $Q$-conservation at the $20 \%$
level or better (e.g., $C1.1$, $C2.1$ and $C3.4$). But it is even
more surprising to find that the end products can be fitted so
well by the $\f$ models. Such good fits make it clear that the
assumption of $Q$ conservation narrows down the very wide range of
self-consistent dynamical models to precisely those few systems
whose properties match both observed systems and the end products
of collisionless collapse. One must conclude that the value of the
$Q$-conservation assumption goes beyond mere ``physical
plausibility" and ``mathematical convenience": it does serve as a
sound physical basis for the construction of dynamical models of
partially relaxed stellar systems.

We should emphasize that such detailed \emph{quantitative}
correspondence with observed systems and with the end products of
collisionless collapse comes as a complete surprise, because the
two parameters that can be varied within the $\f$ family of models
(i.e., $\nu$ and $\Psi$) leave very little freedom with respect to
density and anisotropy profiles (see Paper I). Especially
noteworthy are not only the match of the density profile over nine
orders of magnitude but also the excellent agreement of the
velocity anisotropy profiles between the $\f$ models and several
end products of collapse from clumpy initial conditions (see
Figs.~\ref{fig:C3.5}-\ref{fig:C3.4} and Table~\ref{tab:fit}).

Yet, one cannot claim that the $\f$ models give a fully
satisfactory description of the phase space structure of systems
produced via incomplete violent relaxation. In fact, the
associated $N(E,J^2)$ distribution is characterized by singular
behavior near the origin in the $(E,J^2)$ plane, which is not
present in the end-states of the simulations. In spite of such
discrepancy between models and end-products of the simulations,
the integrated properties (e.g., $N(E)$, $\alpha(r)$ and
$\rho(r)$) are very well reproduced. This confirms the fact that a
variety of different distributions in phase space can lead to the
same integrated properties. In this respect, it appears that, if
we refer to the extreme outer parts of the system (with $r \gg
r_M$, and $E \to 0$) the previously studied $f_{\infty}$ models
\citep{ber84}, with their regular distribution function $f(E,J^2)
\approx |E|^{3/2}$ at low values of $|E|$, might still have an
advantage over the $\f$ models.

Another interesting (although partly known) result of the present
paper is that the velocity distributions of the end products of
the collapse simulations and of the best fitting models possess,
in many cases, a rather strong radial anisotropy. In some of the
collapse simulations we see clear signs that the radial-orbit
instability has been active (as indicated by the correlation
between final ellipticity $\eta$ and anisotropy content
$2K_r/K_T$; cf. Fig.~\ref{fig:corr}), resulting in end products
that are close to the threshold for the onset of the radial-orbit
instability. In general, systems that are unstable with respect to
the radial-orbit instability should evolve into marginally stable
systems (see also the study of the unstable $(1;3.2)$ $\f$ model
in Paper I). In view of the good correspondence between
the results of the formation processes studied in this paper and
important observed properties of elliptical galaxies, we may argue
that ellipticals are also likely to lie close to the threshold of
radial-orbit instability. This would happen if elliptical
galaxies, during their formation process, indeed went through a
collisionless phase characterized by strong radial motions (such
as collapse or head-on mergers). We plan to better quantify this
connection by extending the study to two-component models and
collapses, also starting from a power spectrum of perturbations
representative of cosmological initial conditions.

The last remark brings us naturally to one final comment. We recall
that, since collisionless dynamics is scale-free, the results obtained
here can also be interpreted as relevant to the description of the
collapse of dark matter halos. Clearly, since we do not include the
effects related to the general Hubble expansion and we do not
initialize our clumpy conditions in terms of the power spectrum of
perturbations appropriate for a given cosmological epoch, a direct
comparison between our set of numerical experiments and the profiles
of dark matter halos obtained in $\Lambda CDM$ simulations
\citep{nfw97,moo98} would not be justified. Still, our experiments can
be considered as one example of final equilibrium realizations of a
dark halo, when initial conditions are varied outside the
prescriptions consistent with the currently accepted cosmological
framework (see also \citealt{lem95}). If we now go back to our
interpretation in terms of the $\f$ models, it is noteworthy to point
out that, although the density profile of the $\f$ models falls off as
$1/r^4$ at large radii, in the inner parts that might correspond to
the regions inside the virial radius (for a definition see
\citealt{nfw97}), the density goes approximately as $1/r^{3.2}$ (see
Sect.~3.1 in Paper I), that is very close to the reported $1/r^3$
value for cosmological simulations \citep{nfw97,moo98}. Since the
outskirts of dark matter halos are ``still collapsing'', and thus
their dynamical conditions are different from those under which we
derived the $\f$ models, this agreement appears surprisingly good and
suggests further investigations.

\begin{acknowledgements}
We would like to thank Luca Ciotti, for a number of useful comments and
suggestions, and Peter Teuben, for his kind help and advice on the
use of the NEMO package. MT acknowledges the hospitality of the
Kapteyn Astronomical Institute of Groningen (NL), where part of this
work has been carried out.
\end{acknowledgements}

\appendix

\section{A quantitative measure of clumpiness}

In order to characterize the degree of clumpiness present in the
initial conditions of our simulations, we may consider, in the
$6$-dimensional phase space, the ratio $cl= \langle
\rho_{local}^{(6)}\rangle /\langle \rho^{(6)}\rangle $ of the mean
local density around particles to the mean density.

We estimate the mean $6$-dimensional density in phase space
$\langle \rho^{(6)}\rangle$ by dividing the number of particles
$N$ by the typical total volume occupied. Since the large-scale
structure in phase space is that of a sphere both in position and
velocity space separately, we compute the total volume as the
product of these two volumes. Each volume is calculated by
assuming that the radius of each sphere is equal to the mean
distance between two randomly chosen particles in the relevant
space (position and velocity respectively); for example, for a
homogeneous density distribution inside a sphere of unit radius,
the radius determined from the adopted procedure would be $\approx
1.03$.

The local density $\rho_{local}^{(6)}$ (required for calculating the
average used in the definition of $cl$) is computed by considering one
particle and by counting the number of neighboring particles
$N_{local}$ within a six-dimensional small sphere of fixed radius
$r_s$ (and thus by assuming an equally weighted norm in the phase
space for positions and velocities). The scale $r_s$ is chosen in
such a way that, on average, a small fixed fraction of the total
number of particles is enclosed. We set this fraction to be $\xi=
\langle N_{local} \rangle /N \approx 1/250$. This choice ensures that
we have, on average, a high filling factor within the small sphere, so
that the effects of biases in the local density estimation, arising
from the coincidence of the center of the local sphere with the
coordinates of a particle, are unimportant (for a discussion on the
construction of unbiased estimators for the local density, see also
\citealt{cas85}).

\begin{figure}
\resizebox{\hsize}{!}{\includegraphics{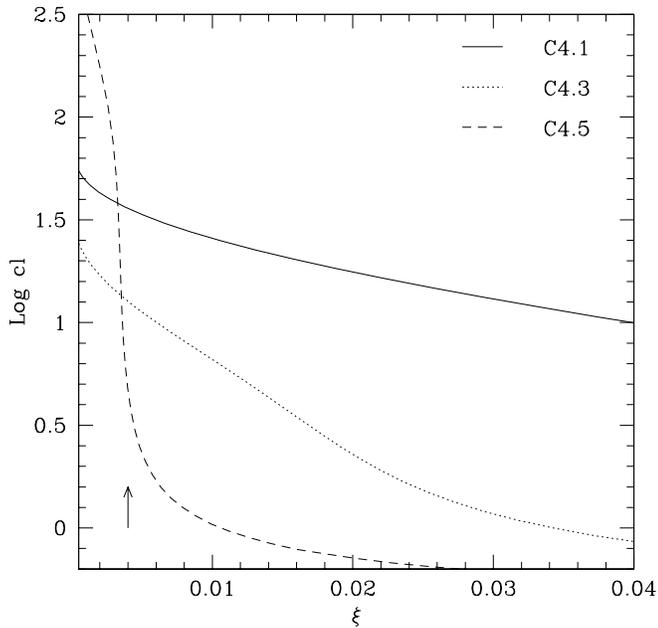}}
\caption{Clumpiness estimator $cl$ as a function of
  $\xi = \langle N_{local} \rangle /N$, for the initial
  conditions of simulations $C4.1$ ($10$ clumps),
  $C4.3$ ($80$ clumps), and $C4.5$ ($400$ clumps). The spatial
  filling factor is kept approximately constant ($N_C \times R_C^3/R^3 = 1.1 - 1.3$). The arrow
  indicates the scale $\xi = 1/250$ at which we refer most of our estimates.}
\label{fig:cl_spectrum}
\end{figure}

The adopted scale $r_s$ also acts as a cut-off scale to the
clumpiness estimator $cl$, which is obviously insensitive to
fluctuations at scales smaller than $r_s$. The dependence of the
clumpiness estimator on $\xi$ is illustrated in
Fig.~\ref{fig:cl_spectrum}. Eventually, diagnostic tools such as
$cl(\xi)$, as a measure of the initial spectrum of inhomogeneities
in phase space, will help us establish a bridge toward initial
conditions representative of the cosmological context (see also
comments at the end of Sect.~\ref{sec:ic_cosmo}).

For our homogeneous initial conditions (simulations of type $U$) the
value of the clumpiness estimator is $0.65 \lesssim cl \lesssim 1$,
depending on the scale considered ($cl=0.72$ for $\xi=1/250$). Note
that the value of $cl$ can fall below unity, because of boundary
effects. In contrast, for the cold clumpy initial conditions of type
$C1$, $C2$, and $C3$ (with $10$ and $20$ clumps, and spatial filling
factor $N_C \times R_C^3/R^3 \approx 1.25$), at $\xi=1/250$ $cl$ takes
on values above $30$, with typical values around $50$ and peaks up to
$100$. For simulation $C4.4$ (with ``small" clumps, and spatial
filling factor $N_C \times R_C^3/R^3 = 0.027$), $cl$ increases to
$300$.  Conversely, $cl$ decreases by increasing the number of clumps
(down to $cl = 15$ for simulation $C4.3$ with $80$ clumps and to $cl
\approx 4.5$ for simulation $C4.5$ with $400$ clumps). 

With the numbers quoted above, we see that, at fixed number of
particles, the clumpiness estimator $cl$ varies with the number of
clumps $N_C$ used.

\subsection{Clumpiness and mixing} \label{sec:cl_mix}

\begin{figure}
\resizebox{\hsize}{!}{\includegraphics{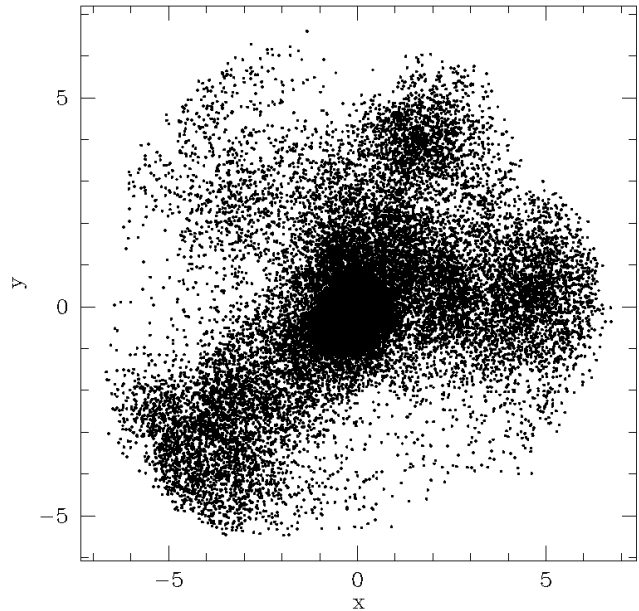}}
\caption{Spatial configuration at time $t = 4$ (i.e., after a few
dynamical times, in the post-collapse phase) for the
  simulation $CV5.1$. Note the presence of clumps in position space.}
\label{fig:posCLvel}
\end{figure}

As already anticipated in Sect.~\ref{sec:jscatter} and
\ref{sec:clumps}, for an efficient angular momentum mixing it is
sufficient that clumpiness be present either in position or velocity
space. In fact, a simulation starting from uniform conditions in terms
of positions but with clumpy structure in velocity space is bound to
develop, after a few dynamical times, a significant clumpiness in
position space (see Fig.~\ref{fig:posCLvel}), so that the
single-particle angular momenta are well mixed at the end of the
simulation (much like in the left panel of Fig.~\ref{fig:jscatter}). This result
confirms that our choice for quantifying the clumpiness of a given
configuration by looking at the six-dimensional phase space is indeed
reasonable.

\bibliographystyle{aa}
\bibliography{PaperTwo}
\end{document}